\documentclass[aps,twocolumn,reprint]{revtex4-2}

\usepackage[utf8]{inputenc}

\usepackage[normalem]{ulem}
\usepackage{color}
\usepackage{mathtools}
\usepackage{graphicx}
\usepackage{hyperref}
\usepackage{xfrac}

\usepackage{comment}

\usepackage{cleveref}
\usepackage{float}
\restylefloat{table}

\usepackage{amsmath,amssymb,amsxtra,amstext,amsfonts,dsfont}
\usepackage{multirow}
\usepackage{array}
\usepackage{upgreek}
\usepackage{mathrsfs}
\usepackage{bbold}
\usepackage{bm}
\usepackage{booktabs}

\usepackage{siunitx} 
\usepackage{soul,xcolor}
\setstcolor{red}

\usepackage{physics}

\newcommand{\oa}{\omega_a}
\newcommand{\oq}{\omega_q}
\newcommand{\om}{\omega}
\newcommand{\Om}{\Omega}

\newcommand{\ad}{{a^\dagger}}
\newcommand{\ada}{a^\dagger a}

\newcommand{\D}{\Delta}

\newcommand{\Hqd}{H_{\text{qd}}}

\newcommand{\Sig}{\Sigma_{21}}
\newcommand{\qdq}{{q}^\dagger q}
\newcommand{\qd}{{q^\dagger}}
\newcommand{\hc}{\mathrm{h.c.}}
\newcommand\mr{\mathrm}
\renewcommand{\var}[1]{\ensuremath{\text{Var}[#1]}}
\newcommand{\cov}[2]{\ensuremath{\, \text{Cov}[#1,#2] }}
\newcommand{\avg}[1]{\left\langle #1 \right\rangle}		

\newcommand{\etal}{\textit{et al.} }

\newcommand{\ga}{g_a}
\newcommand{\gb}{g_b}
\newcommand{\ob}{\omega_b}
\newcommand{\bd}{{b^\dagger}}
\newcommand{\bdb}{b^\dagger b}

\newcommand{\gtms}{g_{\text{TMS}}}
\newcommand{\tsq}{t_{\text{TMS}}}
\newcommand{\trpn}{t_{\text{RPN}}}

\newcommand*{\hbaraux}[2]{\sbox0{\mathsurround=0pt$#1\mathchar'26$}\mkern-1mu\lower.07\ht0\box0\mkern-8mu}

\begin{document}

\title{Einstein--Podolsky--Rosen correlations between mechanical oscillators revealed through SU(1,1) interferometry}

\author{Max-Emanuel Kern}
\affiliation{Department of Physics, ETH Z\"{u}rich, 8093 Z\"{u}rich, Switzerland}

\author{Stefano Marti}
\affiliation{Department of Physics, ETH Z\"{u}rich, 8093 Z\"{u}rich, Switzerland}

\author{Raquel Garcia-Belles}
\affiliation{Department of Physics, ETH Z\"{u}rich, 8093 Z\"{u}rich, Switzerland}

\author{Andraž Omahen}
\affiliation{Department of Physics, ETH Z\"{u}rich, 8093 Z\"{u}rich, Switzerland}

\author{Igor Kladari\'{c}}
\affiliation{Department of Physics, ETH Z\"{u}rich, 8093 Z\"{u}rich, Switzerland}

\author{Arianne Brooks}
\affiliation{Department of Physics, ETH Z\"{u}rich, 8093 Z\"{u}rich, Switzerland}

\author{Yiwen Chu}
\affiliation{Department of Physics, ETH Z\"{u}rich, 8093 Z\"{u}rich, Switzerland}

\author{Matteo Fadel}
\email{fadelm@phys.ethz.ch}
\affiliation{Department of Physics, ETH Z\"{u}rich, 8093 Z\"{u}rich, Switzerland}

\date{\today}
 
\begin{abstract}
Quantum correlations are essential for achieving quantum advantage in computing, communication and sensing. Moreover, their observation challenges and constrains our fundamental understanding of nature. 
Mechanical oscillators in the quantum regime provide an appealing platform for preparing and investigating quantum correlations at macroscopic scales. 
Despite substantial progress, however, continuous-variable quantum correlations stronger than entanglement have not yet been observed in this macroscopic regime.
Here, we report the experimental observation of continuous-variable Einstein--Podolsky--Rosen correlations between two spatially-separated mechanical oscillators with an effective mass of $\sim\SI{16}{\micro\gram}$ each. 
This is achieved by coupling them to a superconducting qubit which allows for engineering a two-mode squeezing interaction when parametrically driven.  
Crucially, we show that this interaction can be used to witness quantum correlations through the realization of a mechanical SU(1,1) interferometer. 
Our results expand the toolbox of operations in circuit quantum acoustodynamics and demonstrate that quantum correlations stronger than entanglement can also be observed in macroscopic systems, thereby shedding light on the boundary between quantum and classical regimes.
\end{abstract}

\maketitle

\begin{figure*}[t]
    \centering
    \includegraphics[width=180mm]{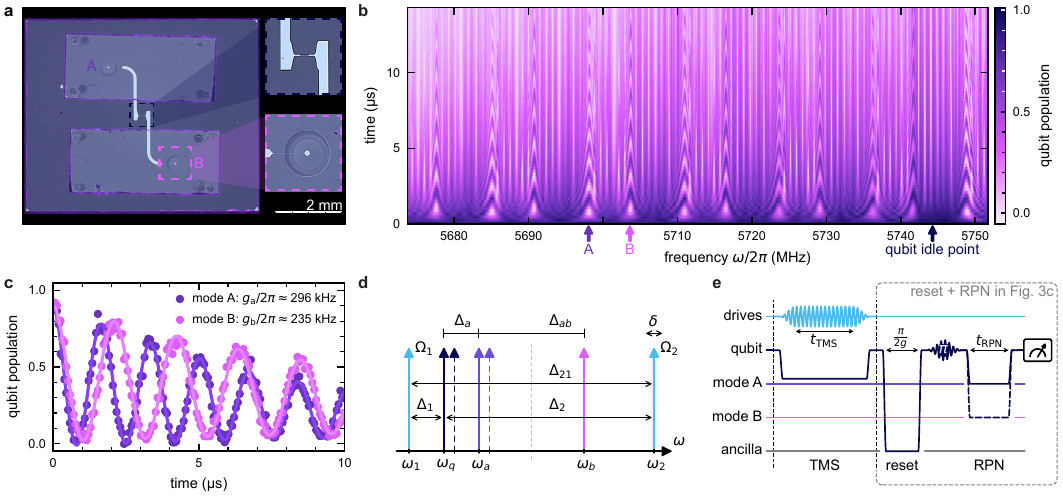}
    \caption{\textbf{
    Device and two-mode squeezing engineering.}
    \textbf{(a)} Micrograph of the device where two HBARs (AlN on sapphire) are bonded on top of a transmon qubit (Al on sapphire). The two elongated capacitive pads of the qubit terminate with antennas designed to optimize the coupling. Insets show detailed views of the Josephson junction and one antenna-HBAR region.
    \textbf{(b)} Spectroscopy of the HBARs' phonon modes measured through vacuum Rabi oscillations versus qubit frequency. The chevron patterns show that the two HBARs have two distinct spectra with non-overlapping modes, allowing us to select the modes A and B that are localized in the two spatially-separated resonators.
    \textbf{(c)} Vacuum Rabi oscillations with the qubit on resonance with mode A or B, from which the qubit-phonon coupling strengths $g_a$ and $g_b$ are extracted, respectively.
    \textbf{(d)} Schematic of the frequency spectrum used for parametric two-mode squeezing. To meet the TMS condition $\omega_1+\omega_2=\omega_a+\omega_b$, a correction $\delta$ to $\omega_2$ is required to account for the drive-induced AC Stark shift of the qubit frequency, which subsequently shifts the frequency of the nearby phonon (dashed lines).
    \textbf{(e)} Pulse sequence used in the TMS rate calibration experiments. After applying the parametric drives to the system for time $\tsq$, we reset the qubit and measure the phonon mode population through a resonant phonon number (RPN) measurement, see main text and Fig.~\ref{fig2}a,b for details.}
    \label{fig1}
\end{figure*}

Quantum theory predicts correlations that defy any classical explanation. Among these, Einstein--Podolsky--Rosen (EPR) steering, first identified by Schr\"{o}dinger~\cite{schrodinger_discussion_1935} in response to the EPR argument on the apparent incompleteness of quantum mechanics~\cite{einstein_can_1935}, describes a form of bipartite nonclassical correlation where a measurement on one system allows one to predict the outcome of a measurement on another system with an accuracy that appears to violate the local uncertainty relation~\cite{reid_demonstration_1989}. 
Importantly, steering is strictly stronger than entanglement and asymmetric under exchange of the two parties~\cite{wiseman_steering_2007}. Such distinctive features make it a resource for quantum information tasks where only one party is assumed to follow the laws of quantum mechanics and the other is not. These include one-sided device-independent quantum key distribution~\cite{branciard_one-sided_2012}, quantum state teleportation~\cite{he_secure_2015}, quantum secret sharing~\cite{he_genuine_2013}, channel discrimination~\cite{piani_necessary_2015}, and assisted quantum metrology~\cite{yadin_metrological_2021}.
Besides its practical applications, the observation of EPR steering provides fundamental insights into the physical world, since it implies that no local hidden-state model description exists~\cite{wiseman_steering_2007}.

Over the past decades, EPR steering has been extensively studied theoretically and demonstrated experimentally across a range of physical platforms~\cite{reid_colloquium_2009,uola_quantum_2020,xiang_quantum_2022}, including optical fields~\cite{ou_realization_1992,handchen_observation_2012} and collective spin variables of atomic ensembles with up to $\sim 10^4$ atoms~\cite{fadel_spatial_2018,kunkel_spatially_2018}.
For more macroscopic systems, however, despite several demonstrations of continuous-variable (CV) entanglement between spatially-separated nanomechanical oscillators~\cite{ockeloen-korppi_stabilized_2018,kotler_direct_2021,mercier_de_lepinay_quantum_2021}, EPR steering has yet to be observed.
This reflects the more stringent requirements on the strength of the prepared quantum correlations and coherence that must be preserved in the system.
Compared to the discrete-variable case, where besides mechanical entanglement~\cite{chou_deterministic_2025}, also Bell correlations have been detected~\cite{MarinkovicPRL}, CV quantum correlations enable different protocols, such as bosonic quantum information processing~\cite{BraunsteinRMP,knill_scheme_2001} and the teleportation of CV states~\cite{bennett_teleporting_1993,furusawa_unconditional_1998}.

Here, we report the experimental observation of continuous-variable EPR correlations between two spatially-separated mechanical oscillators. 
Our platform employs two high-overtone bulk acoustic wave resonators (HBARs) and couples them to a superconducting transmon qubit.
By parametrically driving the system with two microwave tones at well-defined frequencies, we are able to engineer an effective two-mode squeezing (TMS) interaction between phonon modes of the HBARs. 
In contrast to its single-mode counterpart~\cite{marti_quantum_2024}, mechanical TMS results in the generation of strongly-correlated phonon pairs across two distinct modes~\cite{ockeloen-korppi_stabilized_2018,kotler_direct_2021,metzner_two-mode_2024,millican_engineering_2025}, key to many applications.
We further exploit our control of the TMS amplitude and phase to demonstrate a mechanical SU(1,1) interferometer, the analogue of a Mach-Zehnder interferometer where beam splitters are replaced by TMS operations. 
Remarkably, by linking the output signal of such an interferometer to a steering criterion, we are able to witness EPR correlations between mechanical degrees of freedom.

Our results enrich the repertoire for controlling phonon modes of HBAR devices in the quantum regime and establish them as a viable platform for CV quantum information protocols that require entanglement and EPR steering. In addition, they shed light on the validity of quantum mechanics in the macroscopic regime.

\vspace{2mm}
\textbf{Device and model.---}
For our experiments, we have designed and fabricated a circuit quantum acoustodynamics (cQAD) device where two HBARs separated by $\sim\SI{3.6}{\milli\meter}$ are coupled to a single superconducting transmon qubit, see Fig.~\ref{fig1}a. 
Each HBAR is fabricated using piezoelectric aluminum nitride on a sapphire substrate~\cite{catSCI23}, giving an acoustic free spectral range (FSR) of $\sim\SI{12.7}{\mega\hertz}$.
As their thicknesses differ by $\sim\SI{0.5}{\micro m}$, their FSRs are incommensurate and result in non-overlapping phonon modes, see Fig.~\ref{fig1}b.
This allows us to resolve modes that are localized in either of the HBARs and to independently address them.

\begin{figure*}[!t]
    \centering
    \includegraphics[width=180mm]{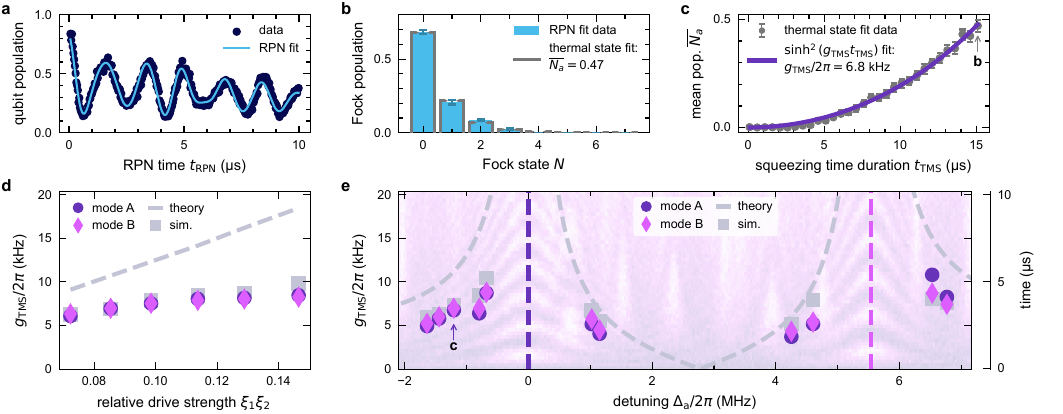}
    \caption{\textbf{TMS rate characterization.} \textbf{(a)} Typical qubit population dynamics when on resonance with the mode to be measured (mode A here) for a time $\trpn$, see Fig.~\ref{fig1}e. The numerical RPN fit extracts the Fock state distribution. 
    \textbf{(b)} Fock state distribution obtained from the data in panel \textbf{a}, with a fit to a thermal state distribution to extract the mean phonon number $\overline{N_a}$.
    \textbf{(c)} Mean population over time $\overline{N_a}(\tsq)$, extracted by repeating the procedure of panels \textbf{a, b} for different TMS durations $\tsq$, with a fit to extract the rate. The arrow indicates the data point from panel \textbf{b}.
    \textbf{(d)} TMS rate $\gtms$, extracted as shown in panels \textbf{a-c}, as a function of the dimensionless parametric drive strength $\xi_1\xi_2$ at fixed detuning $\Delta_a=\SI{-1.224(6)}{\mega\hertz}$.
    Theoretical prediction given by Eq.~\eqref{eq:gTMSanalytic} shows a deviation likely due to effects of higher order in $\xi_1\xi_2$ not included in the theory~\cite{BSpaper,marti_quantum_2024}. Time-dependent simulations show better agreement with the data.
    \textbf{(e)} Same as \textbf{d} but as a function of the qubit-phonon detuning $\Delta_a$ with fixed $\xi_1\xi_2=0.088(4)$. The arrow indicates data from \textbf{a-c} for mode A.
    The background data (see Fig.~\ref{fig1}b) highlight the location of higher-order modes of the HBARs (secondary y-axis on the right).
    }
    \label{fig2}
\end{figure*}

At its idle point, the qubit has a frequency $\omega_{q, \text{idle}}/2\pi = \SI{5.744}{\giga\hertz}$, energy relaxation time $T_{1,q}=\SI{17.9(1.1)}{\micro\second}$, Ramsey decoherence time $T_{2,q}^*=\SI{20.6(1.2)}{\micro s}$ and anharmonicity $\alpha/2\pi = \SI{113}{\MHz}$.
We can tune the qubit frequency using the AC Stark shift resulting from a far-off-resonant microwave drive, which we use to characterize phonon modes through standard cQAD protocols~\cite{BSpaper}.
We choose to perform experiments with the two modes labeled `A' and `B' in Fig.~\ref{fig1}b, at frequencies $\omega_a/2\pi = \SI{5.698}{\giga\hertz}$ and $\omega_b/2\pi = \SI{5.704}{\giga\hertz}$.
These show energy relaxation times $T_{1,a}=\SI{98(5)}{\micro s}$, $T_{1,b}=\SI{186(9)}{\micro s}$, respectively, and $T_1$-limited decoherence times~\cite{belles_loss_2026}. 
The respective coupling strengths are $g_a/2\pi = \SI{296.4(1.8)}{kHz}$ and $g_b/2\pi = \SI{234.9(1.4)}{kHz}$, see Fig.~\ref{fig1}c.
The Hamiltonian describing our cQAD device reads
\begin{align}
    H/\hbar &= \omega_q \qd q - \dfrac{\alpha}{2} {\qd}^2 q^2 + \omega_a \ad a + \omega_b \bd b \notag\\
    &\; + g_a (q \ad + \qd a) + g_b (q \bd + \qd b) + \Hqd/\hbar \;, \label{eq:fullH}
\end{align}
where $q$ and $a, b$ are the bosonic annihilation operators for the qubit and the phonon modes A and B, respectively. 
The term $H_{\text{qd}}/\hbar=(\Omega_1 e^{-i\omega_1 t} + \Omega_2 e^{-i\omega_2 t - i\phi'} ) q^\dagger+\text{h.c.}$ describes two off-resonant microwave drives at frequencies $\omega_{1,2}$, amplitudes $\Omega_{1,2}$, and relative phase $\phi'$ applied to the qubit for engineering parametric interactions~\cite{BSpaper,marti_quantum_2024}, see Fig.~\ref{fig1}d. 
We define the detunings $\Delta_{1,2}=\omega_{1,2}-\omega_q$, $\Sig=\D_1+\D_2$,
and the dimensionless drive strengths $\xi_{1,2}=|\Omega_{1,2}/\Delta_{1,2}|$. Note that we have excluded the AC Stark shift drive to control the qubit frequency  from the Hamiltonian for simplicity. 

When the frequencies of the two parametric drives fulfill the resonance condition $\omega_1+\omega_2 = \omega_a+\omega_b$, the qubit nonlinearity mediates a four-wave mixing process that results in the creation of entangled phonon pairs distributed between modes A and B. 
The emergence of this two-mode squeezing term can be unveiled through a series of unitary transformations (see~\cite{SM} for details), which result in the effective Hamiltonian
\begin{equation}\label{eq:Htms}
    H/\hbar = \Delta \ad a + \gtms (a^\dagger b^\dagger e ^{-i\phi} + a b e^{i\phi} ) \;.
\end{equation}
Here, $\D\approx\left(2 g_a^2 / \D_a +  \oa + \ob - \omega_1 - \omega_2 \right)/2$, and 
\begin{equation}\label{eq:gTMSanalytic}
    \gtms = 2 \xi_1 \xi_2 g_a g_b \frac{\D_a + \D_b}{\D_a \D_b}  \frac{\alpha}{\Sig + \alpha}\;
\end{equation}
is the TMS rate, with $\D_a=\omega_a - \omega_q^{ss}$ the detuning between the phonon mode and the AC Stark shifted qubit, and $\phi$ is the TMS phase resulting from the relative drive phase $\phi'$ after transforming in the frame of mode B~\cite{SM}.
The TMS interaction given by Eq.~\eqref{eq:Htms} is a cornerstone model of quantum optics and its demonstration in cQAD opens new possibilities for CV operations, sensing, and the study of quantum correlations in macroscopic systems using HBAR devices.

\vspace{2mm}
\textbf{TMS calibration.---} To calibrate the TMS process in our device as a function of the dimensionless parametric drive strength $\xi_1\xi_2$ and qubit-phonon detuning $\Delta_a$, we first need to find the right resonance condition. This is necessary to account for drive-induced frequency shifts and dispersive shifts of the phonon modes. To determine the resonance condition, we sweep the frequency $\omega_2$ of one drive by a small correction $\delta$, as indicated in Fig.~\ref{fig1}d. For each $\omega_2+\delta$, the drives are applied for a fixed time $\tsq$, after which the $\ket{1}$ population is read out separately for each phonon mode via the qubit~\cite{marti_quantum_2024}. The desired resonance condition is then identified from the peaks of the phonon populations, as this corresponds to the activation of a parametric process generating excitations.

\begin{figure*}[!t]
    \centering
    \includegraphics[width=120mm]{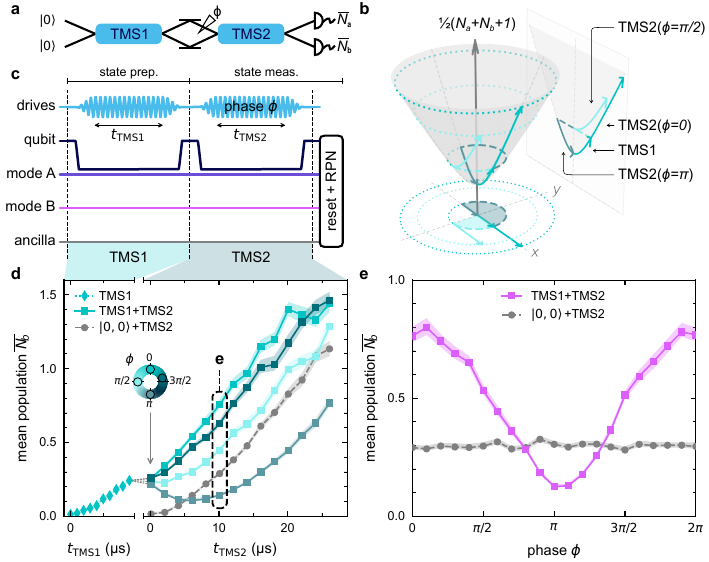}
    \caption{\textbf{Mechanical SU(1,1) interferometry}. \textbf{(a)} Schematic of an SU(1,1) interferometer, where the output is sensitive to the phase $\phi$ accumulated between two TMS operations. \textbf{(b)} Visualization of the SU(1,1) interferometer dynamics on a hyperboloid and projections onto the $xy$- and $xz$-planes. After the initial TMS1 boost along $x$, the phase $\phi$ gives a rotation around $z$. The final TMS2 boost, along $x$, results in an output population that depends on $\phi$. \textbf{(c)} Pulse sequence for the interferometer experiment, similar to Fig.~\ref{fig1}e but including a second TMS operation with a phase $\phi$. \textbf{(d)} Mean population in mode B during TMS1 and TMS2 over time for four different phases $\phi$, as indicated by the inset color wheel. The gray dots correspond to a reference measurement obtained by turning off TMS1, such that TMS2 starts from $\ket{0,0}$. \textbf{(e)} SU(1,1) interferometer fringe measured in mode B for $t_{\text{TMS1}}=\SI{9}{\micro\second}$ and $t_{\text{TMS2}}=\SI{10}{\micro\second}$, which is also used to calibrate the resonance condition for the SU(1,1) interferometer. The gray dots again correspond to the calibration measurement in panel \textbf{d}.}
    \label{fig3}
\end{figure*}

With the resonance condition found, we can now follow a fitting routine to calibrate the TMS strength $\gtms$.
To this end, we need to characterize the state resulting from the two-phonon drive: we fix $\Delta_a$, $\xi_1\xi_2$, and $t_{\rm TMS}$ to the desired values, reset the qubit~\cite{Omahen_reset_2026}, and measure the phonon-number distribution in each mode via resonant phonon number (RPN) measurements~\cite{rahman_genuine_2024,yang_mechanical_2024}. For the RPN measurement, we prepare the qubit in the excited state, let it interact resonantly with the target mode for a time $\trpn$, and then measure its population, following the pulse sequence in Fig.~\ref{fig1}e.
Repeating this measurement for different $\trpn$ and averaging over 
noise results in a trace like the one shown in Fig.~\ref{fig2}a. 
From this data, we reconstruct the Fock-state distribution shown in Fig.~\ref{fig2}b by fitting the data with a set of basis functions. These basis functions are obtained from numerical simulations of the qubit dynamics during resonant interactions with a mode prepared in different Fock states. The weights assigned to each basis function in the fit then correspond to the population of that Fock state $N$.
Formally, this measurement corresponds to measuring the diagonal elements of the reduced density matrix for the mode under investigation.
For ideal TMS dynamics, we expect the reduced state to be a thermal state $\rho_{a}^\text{Th}(\overline{N_a}) = \frac{1}{\overline{N_a}+1} \sum_n \left(\frac{\overline{N_a}}{\overline{N_a}+1}\right)^n \ket{n}\bra{n} $ for mode A, with mean phonon number $\overline{N_a} = \sinh^2 r$, where $r=\gtms\,t_{\rm TMS}$.
As shown in Fig.~\ref{fig2}b, the measured Fock-state distributions closely follow that of a thermal state, allowing us to extract $\overline{N_a}$ by fitting a thermal state (i.e., Bose-Einstein) distribution to the histogram. 
Performing this measurement of the mean phonon number for different squeezing times $t_{\rm TMS}$, we obtain the data presented in Fig.~\ref{fig2}c, showing the expected growth in $\overline{N_a}(t_{\rm TMS})$.
A fit of the analytical model $\sinh^2(\gtms\, t_{\rm TMS})$ then allows us to extract the TMS rate $\gtms$.

Using this fitting routine, we now investigate the dependence of $\gtms$ on the experimentally tunable parameters appearing in Eq.~\eqref{eq:gTMSanalytic}.
For this, we repeat the routine for several combinations of $\Delta_a$, $\xi_1\xi_2$, and measured mode, see Fig.~\ref{fig2}d,e. 
For each combination of parameters, the strengths of the drives are extracted from an independent measurement of their induced AC Stark shift to the qubit. 
The detuning of the qubit and the phonon mode during the TMS process is measured in a spectroscopy measurement and then corrected for their normal mode splitting~\cite{SM}. 
From Fig.~\ref{fig2}d,e, we see that the measured $\gtms$ qualitatively follows the theoretical prediction of Eq.~\eqref{eq:gTMSanalytic}, although showing a deviation that we attribute to effects of higher order in $\xi_1\xi_2$ not included in the theory. 
To support this, we have carried out time-dependent simulations of Eq.~\eqref{eq:fullH} using the values of $\Delta_a$ and $\xi_1\xi_2$ for each point, and the independently characterized qubit/phonon properties~\cite{SM}. We see that the simulation results agree with the experimental $\gtms$ more closely, with larger deviations only for high drive powers, small detunings, and for $0<\Delta_a<\text{FSR}$.

\begin{figure*}[!t]
    \centering
    \includegraphics[width=89mm]{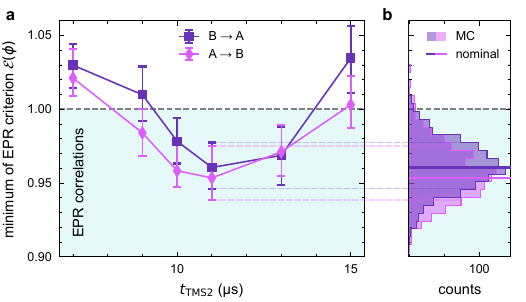}
    \caption{\textbf{Witnessing EPR correlations}. \textbf{(a)} Minimum of the EPR steering criterion $\mathcal{E}(\phi)$ extracted from SU(1,1) interferometer fringes (see Fig.~\ref{fig3}e) for $t_\text{TMS1} = \SI{9}{\micro\second}$ at the optimal phase ($\phi\approx\pi$). \textbf{(b)} Monte Carlo histogram used to determine the error bars of panel \textbf{a}. At the minimum of $t_\text{TMS2} = \SI{11}{\micro\second}$, we find 
    99\% (98\%) of the Monte Carlo samples exhibit EPR correlations for the steering direction B$\rightarrow$A (A$\rightarrow$B).}
    \label{fig4}
\end{figure*}

\vspace{2mm}
\textbf{SU(1,1) interferometry.---}
Thus far, we have characterized the TMS rate, but the presented measurements do not yet demonstrate coherence in the phonon-pair creation process.
For this reason, we now introduce a measurement scheme based on SU(1,1) interferometry which is sensitive to the coherence between the two modes.

In a classical SU(2), i.e. Mach-Zehnder, interferometer, a first beam splitter generates coherence between two modes, and a second beam splitter converts their accumulated relative phase into a measurable population imbalance. 
Crucially, in SU(2) interferometry, the total particle number is conserved, allowing the dynamics to be represented on the surface of a sphere. The system state corresponds to a point whose position encodes the population imbalance and relative phase between the two modes. Beam splitters and phase accumulation act as rotations of this point on the sphere, providing an intuitive geometric picture of the dynamics.

By contrast, an SU(1,1) interferometer replaces the beam splitters with parametric amplifiers, here implemented via TMS interactions, see Fig.~\ref{fig3}a. 
The first TMS (TMS1) generates pairs of coherent excitations in the two modes, which subsequently interfere at the second TMS (TMS2). Because TMS is a phase-sensitive amplifier~\cite{caves_quantum_1982}, its output depends strongly on the coherence between the input modes. Since excitations are always created and annihilated in pairs, the population difference between the two modes in an SU(1,1) interferometer is conserved, allowing the dynamics to be represented on the surface of a hyperboloid~\cite{yurke_su2_1986}, see Fig.~\ref{fig3}b.
Starting from the lower vertex, corresponding to the two-mode vacuum $\ket{0,0}$, TMS1 acts as a boost along the positive $x$-direction. Phase accumulation then induces a rotation about the $z$-axis, which is followed by TMS2. Finally, population measurement corresponds to a projection onto the $z$-axis, yielding interference fringes as a function of the accumulated phase.
For a rigorous mathematical description see~\cite{SM}.

Our implementation of a mechanical SU(1,1) interferometer is illustrated in Fig.~\ref{fig3}c. Compared to the sequence in Fig.~\ref{fig1}e, it includes a second TMS operation after the first one. Note that the duration of this second operation $t_\text{TMS2}$ does not need to be equal to $t_\text{TMS1}$, and that a relative phase $\phi$ between the interferometer arms is introduced by varying the phase of the TMS2 drives compared to the TMS1 drives~\cite{SM}. To better understand how this scheme allows for probing coherence and quantum correlations originating from our engineered TMS interaction, it is instructive to see the TMS1 operation as the state preparation, while viewing TMS2 as part of the readout of this state.

In Fig.~\ref{fig3}d we show the population $\overline{N_b}$ as a function of $t_{\rm TMS1}$ and $t_{\rm TMS2}$, extracted following the same data analysis routine as in Fig.~\ref{fig2}a,b.
Starting from the phonon modes in $\ket{0,0}$, applying only TMS1 results in a fast increase of $\overline{N_b}$, analogously to what is shown in Fig.~\ref{fig2}c.
We then fix $t_{\rm TMS1}=\SI{9}{\micro\second}$ and apply TMS2 for different times $t_{\rm TMS2}$ and drive phases $\phi$.
The resulting output population shows either a further increase or a decrease depending on $\phi$, characteristic 
of a phase-coherent input for TMS2 and the subsequent phase-sensitive amplification process. 
This case can be compared to a reference measurement, where we turn off TMS1 such that the TMS2 operation starts from the phase-invariant $\ket{0,0}$ state, resulting in the $\phi$-independent gray line shown in Fig.~\ref{fig3}d.
The phase sensitivity of our SU(1,1) interferometer becomes even more apparent in Fig.~\ref{fig3}e, where we fix $t_{\rm TMS1}=\SI{9}{\micro\second}$, $t_{\rm TMS2}=\SI{10}{\micro\second}$ and then plot $\overline{N_b}$ as a function of $\phi$ for TMS1 on or off. 
With TMS1 on, we obtain the pink line showing the characteristic interference fringe, while turning TMS1 off results in a constant population corresponding to $\sinh^2(\gtms \, t_{\rm TMS2})$. 
Interestingly, this measurement also gives us an accurate method to calibrate the TMS resonance condition by choosing the drive frequency correction $\delta$ such that the minimum of the fringe is at $\phi\approx\pi$~\cite{SM}. We used this calibration method for all experiments involving two TMS operations.

Importantly, note from Fig.~\ref{fig3}e that the TMS2 output population resulting from a TMS1 input can be lower than the one resulting from vacuum. This observation plays a crucial role for the following detection of quantum correlations.

\vspace{2mm}
\textbf{Witnessing quantum correlations.---}
TMS is a paradigmatic process for the generation of quantum correlations in CV systems, such as entanglement and Einstein--Podolsky--Rosen steering.
Practical criteria for detecting these correlations are commonly formulated in terms of quadrature measurements.
For example, introducing $\Delta_\text{sum} \equiv \text{Var}\left[ X_a + g_x X_b \right] +  \text{Var}\left[ P_a + g_p P_b \right]$ with $\{X_i,P_i\}$ quadrature measurements on system $i$ and $\{g_x,g_p\}$ arbitrary real numbers, the condition $\Delta_\text{sum}<(1+|g_x g_p|)$ is sufficient to reveal entanglement~\cite{giovannetti_characterizing_2003} and $\Delta_\text{sum}<1$ to reveal EPR steering from B to A~\cite{reid_demonstration_1989}.
The latter bound is more demanding, reflecting the fact that EPR steering is a strong form of nonclassical correlations for which entanglement is necessary but not sufficient.

In several experimental platforms, such as circuit QED, cQAD, and trapped ions, reading out CV degrees of freedom through two-level systems makes quadrature observables hard to access.
For this reason, we consider here a different approach, which consists of analyzing the quantum correlations at the output of our TMS operation by applying a second TMS operation before performing a phonon-number measurement, as in the SU(1,1) interferometer sequence shown in Fig.~\ref{fig3}a.
Interestingly, this interferometric approach is related to measurement-after-interaction and (Loschmidt) echo techniques used in metrological protocols~\cite{hosten_quantum_2016,burd_quantum_2019,guo_quantum_2024}, but here applied to the detection of quantum correlations~\cite{guo_detection_2026}.

Specifically, we show that the output population $\langle a_{\text{out}}^\dagger a_{\text{out}} \rangle$ measured after a TMS operation of strength $r_2$ and phase $\phi$ can be related to quadrature measurements performed on its input modes as~\cite{SM}
\begin{align}\label{eq:critE}
    \mathcal{E}(\phi) &\equiv \dfrac{2 \langle a_{\text{out}}^\dagger a_{\text{out}} \rangle + 1}{\cosh^2 r_2} \\
    & \geq\text{Var}\left[ X_a + g_x X_b(\phi) \right] +  \text{Var}\left[ P_a + g_p P_b(\phi) \right],
\end{align}
where $X_b(\phi) = \cos \phi X_b - \sin\phi P_b$, $P_b(\phi) = \sin\phi X_b + \cos\phi P_b$, and $g_x = \tanh r_2 = -g_p$.
Remarkably, this expression relates a population measurement to $\Delta_\text{sum}$, allowing us to state that for all separable states $\mathcal{E} \geq ( 1 + \tanh^2 r_2 )$ and for all non-steerable B$\rightarrow$A states $\mathcal{E} \geq 1$.
Thus, a violation of these inequalities witnesses entanglement or B$\rightarrow$A EPR correlations on the input state, respectively. The opposite steering direction is investigated by simply replacing $a_{\rm out}$ by $b_{\rm out}$ in Eq.~\eqref{eq:critE}.

We investigate EPR steering in our system by analyzing SU(1,1) fringes as in Fig.~\ref{fig3}e for a fixed $t_\text{TMS1}=\SI{9}{\micro\second}$ and different $t_\text{TMS2}$. For each chosen phase $\phi\in[0,2\pi]$, we compute the numerator of the witness Eq.~\eqref{eq:critE} from the population obtained when TMS1 is on and the denominator from the reference measurement obtained when TMS1 is off. 
The minimum measured value $\min_{\phi} \mathcal{E}(\phi)$, typically occurring at $\phi\approx\pi$, is presented in Fig.~\ref{fig4}a as a function of $t_\text{TMS2}$.
From our results, we see an optimal witness violation at $t_\text{TMS2} \approx \SI{11}{\micro\second}$.
We note that for an ideal two-mode squeezed vacuum state generated by a TMS1 operation of strength $r_1$, the optimal TMS2 operation to be applied before detection has strength $r_2=2 r_1$~\cite{SM}.
Because of losses and other experimental imperfections, however, the optimal $r_2$ can be reduced.
The uncertainties on our measurements are estimated through Monte Carlo uncertainty propagation~\cite{rahman_genuine_2024}, as discussed in detail in the SM~\cite{SM}. 
From the resulting Monte Carlo distributions shown in Fig.~\ref{fig4}b, we identify a violation of the B$\rightarrow$A (A$\rightarrow$B) EPR steering criterion with 99\% (98\%) confidence. 
Taking into account that each of the two mechanical modes has an effective mass of approximately $\SI{16}{\micro\gram}$~\cite{catSCI23}, corresponding to $\sim 10^{17}\,$atoms, our results witness EPR correlations between macroscopic systems.

\vspace{2mm}
\textbf{Conclusions.---}
In this work, we have demonstrated an effective two-mode squeezing interaction between gigahertz-frequency phonon modes of two spatially-separated HBAR devices. 
This has allowed us to demonstrate a mechanical SU(1,1) interferometer, with immediate applications in EPR steering-based quantum sensing with mechanical degrees of freedom~\cite{yadin_metrological_2021}.
Crucially, we show that the interferometric signal can be used to witness quantum correlations, which in our case are sufficiently strong to allow for two-way Einstein--Podolsky--Rosen steering.
Such correlations are the essential resource for enabling the quantum teleportation of continuous-variable quantum states of motion between mechanical oscillators.
Additionally, in combination with the previously demonstrated beam splitter~\cite{BSpaper} and single-mode squeezing~\cite{marti_quantum_2024} operations, our results enrich the available cQAD toolbox for CV quantum information processing and bosonic quantum simulation in HBARs. 
This opens up the possibility to use the large number of modes available in these devices for hardware-efficient quantum chemistry simulations~\cite{Wang2019,Chen2023}, as well as for boson sampling with phonons.

\vspace{2mm}
{\centering\textbf{Acknowledgements}\\}
\noindent The authors thank P. Gigon for useful discussions and feedback on the manuscript, and Ines C. Rodrigues, Uwe von L\"upke, and Jonathan Knoll for early contributions to the device design. 
Fabrication of the device was performed at the FIRST cleanroom of ETH Z\"urich and the BRNC cleanroom of IBM Z\"urich. 
MF was supported by the Swiss National Science Foundation Ambizione Grant No. 208886, and The Branco Weiss Fellowship -- Society in Science, administered by the ETH Z\"{u}rich.

\vspace{2mm}
{\centering\textbf{Data and code availability}\\}
\noindent Raw data and analysis scripts will be made available on Zenodo. Additional material is available from the corresponding author on reasonable request.\\

\bibliographystyle{apsrev4-1} 

\let\oldaddcontentsline\addcontentsline
\renewcommand{\addcontentsline}[3]{}
\bibliography{mybib}

@article{SM,
journal = {See Supplementary Material}
}

@article{catSCI23,
author = {Marius Bild  and Matteo Fadel  and Yu Yang  and Uwe von L\"upke  and Phillip Martin  and Alessandro Bruno  and Yiwen Chu },
title = {{S}chr\"odinger cat states of a 16-microgram mechanical oscillator},
journal = {Science},
volume = {380},
number = {6642},
pages = {274-278},
year = {2023},
doi = {10.1126/science.adf7553},
URL = {https://www.science.org/doi/abs/10.1126/science.adf7553}}

@article{BSpaper,
	author = {von L{\"u}pke, Uwe and Rodrigues, Ines C. and Yang, Yu and Fadel, Matteo and Chu, Yiwen},
	date = {2024/01/25},
	id = {von L{\"u}pke2024},
	isbn = {1745-2481},
	journal = {Nature Physics},
    volume={20},
	title = {Engineering multimode interactions in circuit quantum acoustodynamics},
	url = {https://doi.org/10.1038/s41567-023-02377-w},
	year = {2024},
    pages={564--570}}

@article{zhang2019engineering,
  title={{Engineering bilinear mode coupling in circuit QED: Theory and experiment}},
  author={Zhang, Yaxing and Lester, Brian J and Gao, Yvonne Y and Jiang, Liang and Schoelkopf, RJ and Girvin, SM},
  journal={Physical Review A},
  volume={99},
  number={1},
  pages={012314},
  year={2019},
  publisher={APS}, 
  doi={http://dx.doi.org/10.1103/PhysRevA.99.012314}
}

@article{Wang2019,
  title = {Efficient Multiphoton Sampling of Molecular Vibronic Spectra on a Superconducting Bosonic Processor},
  author = {Wang, Christopher S. and Curtis, Jacob C. and Lester, Brian J. and Zhang, Yaxing and Gao, Yvonne Y. and Freeze, Jessica and Batista, Victor S. and Vaccaro, Patrick H. and Chuang, Isaac L. and Frunzio, Luigi and Jiang, Liang and Girvin, S. M. and Schoelkopf, Robert J.},
  journal = {Phys. Rev. X},
  volume = {10},
  issue = {2},
  pages = {021060},
  numpages = {18},
  year = {2020},
  month = {Jun},
  publisher = {American Physical Society},
  doi = {10.1103/PhysRevX.10.021060},
  url = {https://link.aps.org/doi/10.1103/PhysRevX.10.021060}
}

@article{vonLupke22,
	author = {von L{\"u}pke, Uwe and Yang, Yu and Bild, Marius and Michaud, Laurent and Fadel, Matteo and Chu, Yiwen},
	date = {2022/07/01},
	doi = {10.1038/s41567-022-01591-2},
	id = {von L{\"u}pke2022},
	isbn = {1745-2481},
	journal = {Nature Physics},
	number = {7},
	pages = {794--799},
	title = {Parity measurement in the strong dispersive regime of circuit quantum acoustodynamics},
	volume = {18},
	year = {2022}}

@article{BraunsteinRMP,
  title = {Quantum information with continuous variables},
  author = {Braunstein, Samuel L. and van Loock, Peter},
  journal = {Rev. Mod. Phys.},
  volume = {77},
  issue = {2},
  pages = {513--577},
  numpages = {0},
  year = {2005},
  month = {Jun},
  publisher = {American Physical Society},
  doi = {10.1103/RevModPhys.77.513},
  url = {https://link.aps.org/doi/10.1103/RevModPhys.77.513}
}

@article{Chen2023,
	author = {Chen, Wentao and Lu, Yao and Zhang, Shuaining and Zhang, Kuan and Huang, Guanhao and Qiao, Mu and Su, Xiaolu and Zhang, Jialiang and Zhang, Jing-Ning and Banchi, Leonardo and Kim, M. S. and Kim, Kihwan},
	date = {2023/06/01},
	doi = {10.1038/s41567-023-01952-5},
	id = {Chen2023},
	isbn = {1745-2481},
	journal = {Nature Physics},
	number = {6},
	pages = {877--883},
	title = {Scalable and programmable phononic network with trapped ions},
	url = {https://doi.org/10.1038/s41567-023-01952-5},
	volume = {19},
	year = {2023}}

@article{chou_deterministic_2025,
	title = {Deterministic multi-phonon entanglement between two mechanical resonators on separate substrates},
	volume = {16},
	issn = {2041-1723},
	url = {https://www.nature.com/articles/s41467-025-56454-0},
	doi = {10.1038/s41467-025-56454-0},
	language = {en},
	number = {1},
	urldate = {2025-02-12},
	journal = {Nature Communications},
	author = {Chou, Ming-Han and Qiao, Hong and Yan, Haoxiong and Andersson, Gustav and Conner, Christopher R. and Grebel, Joel and Joshi, Yash J. and Miller, Jacob M. and Povey, Rhys G. and Wu, Xuntao and Cleland, Andrew N.},
	month = feb,
	year = {2025},
	pages = {1450},
}

@article{ockeloen-korppi_stabilized_2018,
	title = {Stabilized entanglement of massive mechanical oscillators},
	volume = {556},
	issn = {0028-0836, 1476-4687},
	url = {https://www.nature.com/articles/s41586-018-0038-x},
	doi = {10.1038/s41586-018-0038-x},
	language = {en},
	number = {7702},
	urldate = {2025-02-19},
	journal = {Nature},
	author = {Ockeloen-Korppi, C. F. and Damsk\"agg, E. and Pirkkalainen, J.-M. and Asjad, M. and Clerk, A. A. and Massel, F. and Woolley, M. J. and Sillanp\"a\"a, M. A.},
	month = apr,
	year = {2018},
	pages = {478--482},
}

@article{kotler_direct_2021,
	title = {Direct observation of deterministic macroscopic entanglement},
	volume = {372},
	issn = {0036-8075, 1095-9203},
	url = {https://www.science.org/doi/10.1126/science.abf2998},
	doi = {10.1126/science.abf2998},
	language = {en},
	number = {6542},
	urldate = {2025-02-19},
	journal = {Science},
	author = {Kotler, Shlomi and Peterson, Gabriel A. and Shojaee, Ezad and Lecocq, Florent and Cicak, Katarina and Kwiatkowski, Alex and Geller, Shawn and Glancy, Scott and Knill, Emanuel and Simmonds, Raymond W. and Aumentado, José and Teufel, John D.},
	month = may,
	year = {2021},
	pages = {622--625},
}

@article{mercier_de_lepinay_quantum_2021,
	title = {Quantum mechanics–free subsystem with mechanical oscillators},
	volume = {372},
	issn = {0036-8075, 1095-9203},
	url = {https://www.science.org/doi/10.1126/science.abf5389},
	doi = {10.1126/science.abf5389},
	language = {en},
	number = {6542},
	urldate = {2025-02-19},
	journal = {Science},
	author = {Mercier De Lepinay, Laure and Ockeloen-Korppi, Caspar F. and Woolley, Matthew J. and Sillanp\"a\"a, Mika A.},
	month = may,
	year = {2021},
	pages = {625--629},
}

@article{rahman_genuine_2024,
  title = {Genuine Quantum Non-Gaussianity and Metrological Sensitivity of Fock States Prepared in a Mechanical Resonator},
  author = {Rahman, Q. Rumman and Kladari\ifmmode \acute{c}\else \'{c}\fi{}, Igor and Kern, Max-Emanuel and Lachman, Lukas and Chu, Yiwen and Filip, Radim and Fadel, Matteo},
  journal = {Phys. Rev. Lett.},
  volume = {134},
  issue = {18},
  pages = {180801},
  numpages = {6},
  year = {2025},
  month = {May},
  publisher = {American Physical Society},
  doi = {10.1103/PhysRevLett.134.180801},
  url = {https://link.aps.org/doi/10.1103/PhysRevLett.134.180801}
}

@article{reid_demonstration_1989,
    title = {Demonstration of the {Einstein}-{Podolsky}-{Rosen} paradox using nondegenerate parametric amplification},
    volume = {40},
    url = {https://link.aps.org/doi/10.1103/PhysRevA.40.913},
    doi = {10.1103/PhysRevA.40.913},
    number = {2},
    journal = {Phys. Rev. A},
    author = {Reid, M. D.},
    month = jul,
    year = {1989},
    pages = {913--923},
}

@article{giovannetti_characterizing_2003,
    title = {Characterizing the entanglement of bipartite quantum systems},
    volume = {67},
    copyright = {http://link.aps.org/licenses/aps-default-license},
    issn = {1050-2947, 1094-1622},
    url = {https://link.aps.org/doi/10.1103/PhysRevA.67.022320},
    doi = {10.1103/PhysRevA.67.022320},
    number = {2},
    urldate = {2025-07-07},
    journal = {Physical Review A},
    author = {Giovannetti, Vittorio and Mancini, Stefano and Vitali, David and Tombesi, Paolo},
    month = feb,
    year = {2003},
    pages = {022320},
}

@article{marti_quantum_2024,
    title = {Quantum squeezing in a nonlinear mechanical oscillator},
    volume = {20},
    issn = {1745-2473, 1745-2481},
    url = {https://www.nature.com/articles/s41567-024-02545-6},
    doi = {10.1038/s41567-024-02545-6},
    number = {9},
    urldate = {2024-11-11},
    journal = {Nature Physics},
    author = {Marti, Stefano and von L\"upke, Uwe and Joshi, Om and Yang, Yu and Bild, Marius and Omahen, Andraz and Chu, Yiwen and Fadel, Matteo},
    month = sep,
    year = {2024},
    pages = {1448--1453},
}

@article{guo_quantum_2024,
    title = {Quantum metrology with a squeezed {Kerr} oscillator},
    volume = {109},
    issn = {2469-9926, 2469-9934},
    url = {https://link.aps.org/doi/10.1103/PhysRevA.109.052604},
    doi = {10.1103/PhysRevA.109.052604},
    number = {5},
    urldate = {2024-11-11},
    journal = {Physical Review A},
    author = {Guo, Jiajie and He, Qiongyi and Fadel, Matteo},
    month = may,
    year = {2024},
    pages = {052604},
}

@article{burd_quantum_2019,
    title = {Quantum amplification of mechanical oscillator motion},
    volume = {364},
    issn = {0036-8075, 1095-9203},
    url = {https://www.science.org/doi/10.1126/science.aaw2884},
    doi = {10.1126/science.aaw2884},
    number = {6446},
    urldate = {2024-05-19},
    journal = {Science},
    author = {Burd, S. C. and Srinivas, R. and Bollinger, J. J. and Wilson, A. C. and Wineland, D. J. and Leibfried, D. and Slichter, D. H. and Allcock, D. T. C.},
    month = jun,
    year = {2019},
    pages = {1163--1165},
}

@article{hosten_quantum_2016,
    title = {Quantum phase magnification},
    volume = {352},
    issn = {0036-8075, 1095-9203},
    url = {https://www.science.org/doi/10.1126/science.aaf3397},
    doi = {10.1126/science.aaf3397},
    number = {6293},
    urldate = {2026-01-28},
    journal = {Science},
    author = {Hosten, O. and Krishnakumar, R. and Engelsen, N. J. and Kasevich, M. A.},
    month = jun,
    year = {2016},
    pages = {1552--1555},
}

@article{yadin_metrological_2021,
    title = {Metrological complementarity reveals the {Einstein}-{Podolsky}-{Rosen} paradox},
    volume = {12},
    issn = {2041-1723},
    url = {https://www.nature.com/articles/s41467-021-22353-3},
    doi = {10.1038/s41467-021-22353-3},
    number = {1},
    urldate = {2024-11-11},
    journal = {Nature Communications},
    author = {Yadin, Benjamin and Fadel, Matteo and Gessner, Manuel},
    month = apr,
    year = {2021},
    pages = {2410},
}

@article{reid_colloquium_2009,
    title = {Colloquium: {The} {Einstein}-{Podolsky}-{Rosen} paradox: {From} concepts to applications},
    volume = {81},
    url = {http://link.aps.org/doi/10.1103/RevModPhys.81.1727},
    doi = {10.1103/RevModPhys.81.1727},
    number = {4},
    journal = {Rev. Mod. Phys.},
    publisher = {American Physical Society},
    author = {Reid, M. D. and Drummond, P. D. and Bowen, W. P. and Cavalcanti, E. G. and Lam, P. K. and Bachor, H. A. and Andersen, U. L. and Leuchs, G.},
    month = dec,
    year = {2009},
    pages = {1727--1751},
}

@article{wiseman_steering_2007,
    title = {Steering, {Entanglement}, {Nonlocality}, and the {Einstein}-{Podolsky}-{Rosen} {Paradox}},
    volume = {98},
    url = {https://link.aps.org/doi/10.1103/PhysRevLett.98.140402},
    doi = {10.1103/PhysRevLett.98.140402},
    number = {14},
    journal = {Phys. Rev. Lett.},
    publisher = {American Physical Society},
    author = {Wiseman, H. M. and Jones, S. J. and Doherty, A. C.},
    month = apr,
    year = {2007},
    pages = {140402},
}

@article{ou_realization_1992,
    title = {Realization of the {Einstein}-{Podolsky}-{Rosen} paradox for continuous variables},
    volume = {68},
    copyright = {http://link.aps.org/licenses/aps-default-license},
    issn = {0031-9007},
    url = {https://link.aps.org/doi/10.1103/PhysRevLett.68.3663},
    doi = {10.1103/PhysRevLett.68.3663},
    number = {25},
    urldate = {2026-04-02},
    journal = {Physical Review Letters},
    author = {Ou, Z. Y. and Pereira, S. F. and Kimble, H. J. and Peng, K. C.},
    month = jun,
    year = {1992},
    pages = {3663--3666},
}

@article{fadel_spatial_2018,
    title = {Spatial entanglement patterns and {Einstein}-{Podolsky}-{Rosen} steering in {Bose}-{Einstein} condensates},
    volume = {360},
    issn = {0036-8075, 1095-9203},
    url = {https://www.science.org/doi/10.1126/science.aao1850},
    doi = {10.1126/science.aao1850},
    number = {6387},
    urldate = {2024-11-11},
    journal = {Science},
    author = {Fadel, Matteo and Zibold, Tilman and Décamps, Boris and Treutlein, Philipp},
    month = apr,
    year = {2018},
    pages = {409--413},
}

@article{kunkel_spatially_2018,
    title = {Spatially distributed multipartite entanglement enables {EPR} steering of atomic clouds},
    volume = {360},
    issn = {0036-8075},
    url = {http://science.sciencemag.org/content/360/6387/413},
    doi = {10.1126/science.aao2254},
    number = {6387},
    journal = {Science},
    publisher = {American Association for the Advancement of Science},
    author = {Kunkel, Philipp and Prüfer, Maximilian and Strobel, Helmut and Linnemann, Daniel and Frölian, Anika and Gasenzer, Thomas and Gärttner, Martin and Oberthaler, Markus K.},
    year = {2018},
    pages = {413--416},
}

@article{handchen_observation_2012,
    title = {Observation of one-way {Einstein}–{Podolsky}–{Rosen} steering},
    volume = {6},
    copyright = {http://www.springer.com/tdm},
    issn = {1749-4885, 1749-4893},
    url = {https://www.nature.com/articles/nphoton.2012.202},
    doi = {10.1038/nphoton.2012.202},
    language = {en},
    number = {9},
    urldate = {2026-04-01},
    journal = {Nature Photonics},
    author = {Händchen, Vitus and Eberle, Tobias and Steinlechner, Sebastian and Samblowski, Aiko and Franz, Torsten and Werner, Reinhard F. and Schnabel, Roman},
    month = sep,
    year = {2012},
    pages = {596--599},
}

@article{schrodinger_discussion_1935,
    title = {Discussion of {Probability} {Relations} between {Separated} {Systems}},
    volume = {31},
    copyright = {https://www.cambridge.org/core/terms},
    issn = {0305-0041, 1469-8064},
    url = {https://www.cambridge.org/core/product/identifier/S0305004100013554/type/journal_article},
    doi = {10.1017/S0305004100013554},
    language = {en},
    number = {4},
    urldate = {2026-04-02},
    journal = {Mathematical Proceedings of the Cambridge Philosophical Society},
    author = {Schrödinger, E.},
    month = oct,
    year = {1935},
    pages = {555--563},
}

@article{einstein_can_1935,
    title = {Can {Quantum}-{Mechanical} {Description} of {Physical} {Reality} {Be} {Considered} {Complete}?},
    volume = {47},
    copyright = {https://link.aps.org/licenses/aps-default-license},
    issn = {0031-899X},
    url = {https://link.aps.org/doi/10.1103/PhysRev.47.777},
    doi = {10.1103/PhysRev.47.777},
    language = {en},
    number = {10},
    urldate = {2026-04-02},
    journal = {Physical Review},
    author = {Einstein, A. and Podolsky, B. and Rosen, N.},
    month = may,
    year = {1935},
    pages = {777--780},
}

@article{branciard_one-sided_2012,
    title = {One-sided device-independent quantum key distribution: {Security}, feasibility, and the connection with steering},
    volume = {85},
    copyright = {http://link.aps.org/licenses/aps-default-license},
    issn = {1050-2947, 1094-1622},
    shorttitle = {One-sided device-independent quantum key distribution},
    url = {https://link.aps.org/doi/10.1103/PhysRevA.85.010301},
    doi = {10.1103/PhysRevA.85.010301},
    language = {en},
    number = {1},
    urldate = {2026-04-02},
    journal = {Physical Review A},
    author = {Branciard, Cyril and Cavalcanti, Eric G. and Walborn, Stephen P. and Scarani, Valerio and Wiseman, Howard M.},
    month = jan,
    year = {2012},
    pages = {010301},
}

@article{he_secure_2015,
    title = {Secure {Continuous} {Variable} {Teleportation} and {Einstein}-{Podolsky}-{Rosen} {Steering}},
    volume = {115},
    copyright = {http://link.aps.org/licenses/aps-default-license},
    issn = {0031-9007, 1079-7114},
    url = {https://link.aps.org/doi/10.1103/PhysRevLett.115.180502},
    doi = {10.1103/PhysRevLett.115.180502},
    language = {en},
    number = {18},
    urldate = {2026-04-02},
    journal = {Physical Review Letters},
    author = {He, Qiongyi and Rosales-Zárate, Laura and Adesso, Gerardo and Reid, Margaret D.},
    month = oct,
    year = {2015},
    pages = {180502},
}

@article{he_genuine_2013,
    title = {Genuine {Multipartite} {Einstein}-{Podolsky}-{Rosen} {Steering}},
    volume = {111},
    copyright = {http://link.aps.org/licenses/aps-default-license},
    issn = {0031-9007, 1079-7114},
    url = {https://link.aps.org/doi/10.1103/PhysRevLett.111.250403},
    doi = {10.1103/PhysRevLett.111.250403},
    language = {en},
    number = {25},
    urldate = {2026-04-02},
    journal = {Physical Review Letters},
    author = {He, Q. Y. and Reid, M. D.},
    month = dec,
    year = {2013},
    pages = {250403},
}

@article{piani_necessary_2015,
    title = {Necessary and {Sufficient} {Quantum} {Information} {Characterization} of {Einstein}-{Podolsky}-{Rosen} {Steering}},
    volume = {114},
    copyright = {http://link.aps.org/licenses/aps-default-license},
    issn = {0031-9007, 1079-7114},
    url = {https://link.aps.org/doi/10.1103/PhysRevLett.114.060404},
    doi = {10.1103/PhysRevLett.114.060404},
    language = {en},
    number = {6},
    urldate = {2026-04-02},
    journal = {Physical Review Letters},
    author = {Piani, Marco and Watrous, John},
    month = feb,
    year = {2015},
    pages = {060404},
}

@article{millican_engineering_2025,
    title = {Engineering {Continuous}-{Variable} {Entanglement} in {Mechanical} {Oscillators} with {Optimal} {Control}},
    volume = {135},
    issn = {0031-9007, 1079-7114},
    url = {https://link.aps.org/doi/10.1103/2ntn-qh3t},
    doi = {10.1103/2ntn-qh3t},
    language = {en},
    number = {23},
    urldate = {2026-04-02},
    journal = {Physical Review Letters},
    author = {Millican, Maverick J. and Matsos, Vassili G. and Valahu, Christophe H. and Navickas, Tomas and Bond, Liam J. and Tan, Ting Rei},
    month = dec,
    year = {2025},
    pages = {233604},
}

@article{metzner_two-mode_2024,
    title = {Two-mode squeezing and {SU}(1,1) interferometry with trapped ions},
    volume = {110},
    issn = {2469-9926, 2469-9934},
    url = {https://link.aps.org/doi/10.1103/PhysRevA.110.022613},
    doi = {10.1103/PhysRevA.110.022613},
    language = {en},
    number = {2},
    urldate = {2026-04-02},
    journal = {Physical Review A},
    author = {Metzner, J. and Quinn, A. and Brudney, S. and Moore, I. D. and Burd, S. C. and Wineland, D. J. and Allcock, D. T. C.},
    month = aug,
    year = {2024},
    pages = {022613},
}

@article{caves_quantum_1982,
    title = {Quantum limits on noise in linear amplifiers},
    volume = {26},
    copyright = {http://link.aps.org/licenses/aps-default-license},
    issn = {0556-2821},
    url = {https://link.aps.org/doi/10.1103/PhysRevD.26.1817},
    doi = {10.1103/PhysRevD.26.1817},
    language = {en},
    number = {8},
    urldate = {2026-04-03},
    journal = {Physical Review D},
    author = {Caves, Carlton M.},
    month = oct,
    year = {1982},
    pages = {1817--1839},
}

@article{bennett_teleporting_1993,
    title = {Teleporting an unknown quantum state via dual classical and {Einstein}-{Podolsky}-{Rosen} channels},
    volume = {70},
    url = {https://link.aps.org/doi/10.1103/PhysRevLett.70.1895},
    doi = {10.1103/PhysRevLett.70.1895},
    number = {13},
    journal = {Phys. Rev. Lett.},
    publisher = {American Physical Society},
    author = {Bennett, Charles H. and Brassard, Gilles and Crépeau, Claude and Jozsa, Richard and Peres, Asher and Wootters, William K.},
    month = mar,
    year = {1993},
    pages = {1895--1899},
}

@book{gerry_introductory_2004,
    edition = {1},
    title = {Introductory {Quantum} {Optics}},
    copyright = {https://www.cambridge.org/core/terms},
    isbn = {978-0-521-52735-4 978-0-521-82035-6 978-0-511-79123-9},
    url = {https://www.cambridge.org/core/product/identifier/9780511791239/type/book},
    doi = {10.1017/CBO9780511791239},
    urldate = {2026-04-03},
    publisher = {Cambridge University Press},
    author = {Gerry, Christopher and Knight, Peter},
    month = oct,
    year = {2004},
}

@article{knill_scheme_2001,
    title = {A scheme for efficient quantum computation with linear optics},
    volume = {409},
    copyright = {http://www.springer.com/tdm},
    issn = {0028-0836, 1476-4687},
    url = {https://www.nature.com/articles/35051009},
    doi = {10.1038/35051009},
    language = {en},
    number = {6816},
    urldate = {2026-04-23},
    journal = {Nature},
    author = {Knill, E. and Laflamme, R. and Milburn, G. J.},
    month = jan,
    year = {2001},
    pages = {46--52},
}

@article{furusawa_unconditional_1998,
    title = {Unconditional {Quantum} {Teleportation}},
    volume = {282},
    issn = {0036-8075, 1095-9203},
    url = {https://www.science.org/doi/10.1126/science.282.5389.706},
    doi = {10.1126/science.282.5389.706},
    language = {en},
    number = {5389},
    urldate = {2026-04-23},
    journal = {Science},
    author = {Furusawa, A. and Sørensen, J. L. and Braunstein, S. L. and Fuchs, C. A. and Kimble, H. J. and Polzik, E. S.},
    month = oct,
    year = {1998},
    pages = {706--709},
}

@article{guo_detection_2026,
    title = {Detection of non-{Gaussian} quantum correlations through measurement-after-interaction protocols},
    volume = {113},
    issn = {2469-9926, 2469-9934},
    url = {https://link.aps.org/doi/10.1103/b9d8-dgbf},
    doi = {10.1103/b9d8-dgbf},
    language = {en},
    number = {1},
    urldate = {2026-04-23},
    journal = {Physical Review A},
    author = {Guo, Jiajie and Sun, Feng-Xiao and Fadel, Matteo and He, Qiongyi},
    month = jan,
    year = {2026},
    pages = {012409},
}

@article{uola_quantum_2020,
    title = {Quantum steering},
    volume = {92},
    issn = {0034-6861, 1539-0756},
    url = {https://link.aps.org/doi/10.1103/RevModPhys.92.015001},
    doi = {10.1103/RevModPhys.92.015001},
    language = {en},
    number = {1},
    urldate = {2026-04-23},
    journal = {Reviews of Modern Physics},
    author = {Uola, Roope and Costa, Ana C. S. and Nguyen, H. Chau and Gühne, Otfried},
    month = mar,
    year = {2020},
    pages = {015001},
}

@article{xiang_quantum_2022,
    title = {Quantum {Steering}: {Practical} {Challenges} and {Future} {Directions}},
    volume = {3},
    issn = {2691-3399},
    shorttitle = {Quantum {Steering}},
    url = {https://link.aps.org/doi/10.1103/PRXQuantum.3.030102},
    doi = {10.1103/PRXQuantum.3.030102},
    language = {en},
    number = {3},
    urldate = {2026-04-23},
    journal = {PRX Quantum},
    author = {Xiang, Yu and Cheng, Shuming and Gong, Qihuang and Ficek, Zbigniew and He, Qiongyi},
    month = aug,
    year = {2022},
    pages = {030102},
}

@misc{belles_loss_2026,
    title = {Loss {Mechanisms} in {High}-coherence {Multimode} {Mechanical} {Resonators} {Coupled} to {Superconducting} {Circuits}},
    url = {http://arxiv.org/abs/2602.22117},
    doi = {10.48550/arXiv.2602.22117},
    urldate = {2026-02-26},
    publisher = {arXiv},
    author = {Belles, Raquel Garcia and Anferov, Alexander and Deeg, Lukas F. and Colicchio, Loris and Brooks, Arianne and Schatteburg, Tom and Drimmer, Maxwell and Rodrigues, Ines C. and Benevides, Rodrigo and Liffredo, Marco and Patidar, Jyotish and Pshyk, Oleksandr and Fadel, Matteo and Villanueva, Luis Guillermo and Siol, Sebastian and Kirchmair, Gerhard and Chu, Yiwen},
    month = feb,
    year = {2026},
    note = {arXiv:2602.22117 [quant-ph]},
}

@misc{Omahen_reset_2026,
      title={High-Fidelity Transmon Reset with a Multimode Acoustic Resonator}, 
      author={Andraž Omahen and Simon Storz and Igor Kladarić and Yiwen Chu},
      year={2026},
      publisher={arXiv},
      url={https://arxiv.org/abs/2604.08655},
      note = {arXiv:2604.08655 [quant-ph]},
}

@article{yurke_su2_1986,
    title = {{SU}(2) and {SU}(1,1) interferometers},
    volume = {33},
    copyright = {http://link.aps.org/licenses/aps-default-license},
    issn = {0556-2791},
    url = {https://link.aps.org/doi/10.1103/PhysRevA.33.4033},
    doi = {10.1103/PhysRevA.33.4033},
    language = {en},
    number = {6},
    urldate = {2024-05-07},
    journal = {Physical Review A},
    author = {Yurke, Bernard and McCall, Samuel L. and Klauder, John R.},
    month = jun,
    year = {1986},
    pages = {4033--4054},
}

@article{MarinkovicPRL,
  title = {Optomechanical Bell Test},
  author = {Marinkovi\ifmmode \acute{c}\else \'{c}\fi{}, Igor and Wallucks, Andreas and Riedinger, Ralf and Hong, Sungkun and Aspelmeyer, Markus and Gr\"oblacher, Simon},
  journal = {Phys. Rev. Lett.},
  volume = {121},
  issue = {22},
  pages = {220404},
  numpages = {6},
  year = {2018},
  month = {Nov},
  publisher = {American Physical Society},
  doi = {10.1103/PhysRevLett.121.220404},
  url = {https://link.aps.org/doi/10.1103/PhysRevLett.121.220404}
}

@article{yang_mechanical_2024,
    title = {A mechanical qubit},
    volume = {386},
    issn = {0036-8075, 1095-9203},
    url = {https://www.science.org/doi/10.1126/science.adr2464},
    doi = {10.1126/science.adr2464},
    language = {en},
    number = {6723},
    urldate = {2024-11-25},
    journal = {Science},
    author = {Yang, Yu and Kladarić, Igor and Drimmer, Maxwell and von Lüpke, Uwe and Lenterman, Daan and Bus, Joost and Marti, Stefano and Fadel, Matteo and Chu, Yiwen},
    month = nov,
    year = {2024},
    pages = {783--788},
}
\let\addcontentsline\oldaddcontentsline

\clearpage
\newpage

\renewcommand{\thetable}{S\arabic{table}}  
\renewcommand{\thepage}{S\arabic{page}}  
\renewcommand{\thefigure}{S\arabic{figure}}
\renewcommand{\theHfigure}{S\arabic{figure}}
\renewcommand{\theequation}{S\arabic{equation}}
\setcounter{page}{1}
\setcounter{figure}{0}
\setcounter{table}{0}
\setcounter{section}{0}
\setcounter{equation}{0}

\widetext

{\centering\textbf{\Large Supplementary Material} \\} 
\normalsize
\vspace{.3cm}

{\centering Max-Emanuel Kern, Stefano Marti, Raquel Garcia-Belles, Andraž Omahen,\\ Igor Kladari\'{c}, Arianne Brooks, Yiwen Chu and Matteo Fadel$^\ast$\\
\textit{Department of Physics, ETH Z\"{u}rich, 8093 Z\"{u}rich, Switzerland \\
$^\ast$ Corresponding author: fadelm@phys.ethz.ch}\\
}

\suppressfloats

\tableofcontents

\clearpage
\newpage

\section{Device and experiment parameters}\label{sec:device_params}

The device is fabricated in an analogous manner to our previous work~\cite{vonLupke22, catSCI23, BSpaper}. In particular, we flip-chip-bonded two plano-convex HBARs (aluminum nitride on sapphire) to a superconducting transmon qubit (aluminum on sapphire) with a single junction and two antennas to fabricate this cQAD device. The coupling between the mechanics and the qubit is mediated by the electric field emitted by the two qubit antennas and the piezoelectricity of the aluminum nitride. The device is mounted inside a 3D superconducting cavity made of aluminum, which is used to shield the device and read out its state. The device is further shielded with a Mu-Metal can and mounted on the base stage of a dilution refrigerator. Further details about the measurement setup can be found in our earlier works, for example the supplementary material of Ref.~\cite{BSpaper}.  

The device parameters are summarized in Table~\ref{tab:system_params}. Note the lower anharmonicity compared to previous devices resulting from the new geometry. Also, the two coupling strengths are unequal, which we attribute to small differences in the distance between the qubit and the respective HBAR.

\begin{table}[H]
    \centering
    \caption{\textbf{Qubit and phonon properties.} The nominal values and the uncertainties of coupling strengths, energy relaxation times, and dephasing times are derived from 11 device characterization measurements taken close to the measurements presented in this work carried out over the course of two months.}
    \label{tab:system_params}
    {\setlength{\tabcolsep}{12pt}
    \begin{tabular}{@{} l c c c c @{}}
        \toprule
        parameter & qubit & mode A & mode B & unit\\ 
        \midrule
        resonance frequency ($\omega/2\pi$) & 5.744 & 5.698
        &
        5.704
        & \unit{\giga\hertz}\\
        anharmonicity ($\alpha/2\pi$) & 112.8 & -- & -- & \unit{{\mega\hertz}}\\
        inter-mode FSR ($\Delta_\text{ab}/2\pi$) & -- & \multicolumn{2}{c}{5.5} & \unit{\mega\hertz} \\
        coupling strength ($g/2\pi$) & -- & 296.4(1.8) & 234.9(1.4) & \unit{\kilo\hertz} \\
        energy relaxation time ($T_1$) & 17.9(1.1) & 98(5) & 186(9) & \unit{\micro\second} \\
        Ramsey decoherence time ($T_2^*$) & 20.6(1.2) & 191(5) & 368(19) & \unit{\micro\second} \\
        \bottomrule
    \end{tabular}}
\end{table}

\clearpage
\newpage

\vspace{10mm}
\section{Measurement Uncertainties and Data Analysis}\label{sec:errorbars}
\hfill\\

Uncertainties in device parameters like qubit idle point and phonon $T_1$ and $T_2^*$, and the coupling strengths $g$ are calculated from statistics of 11 device characterization measurements taken before the experiments presented here carried out over the course of two months. The uncertainty is low for the coupling strength $g$ in particular.

Uncertainties on qubit population traces are on the scale of the marker size and are obtained from binomial shot noise statistics with $n_\text{avg}=1000$ for Fig.~\ref{fig1}c and $n_\text{avg}=2000$ for Fig.~\ref{fig2}a. For all other RPN measurements we also took $n_\text{avg}=2000$ measurements.

Because of the intricate fitting routine, we model the uncertainties in all other measurements via Monte Carlo (MC) uncertainty propagation~\cite{rahman_genuine_2024}, i.e., in Fig.~\ref{fig2}, Fig.~\ref{fig3}d,e, and Fig.~\ref{fig4}. For each of the four possible final parameters $N, \overline{N_{a,b}}, \gtms$ or $\mathcal{E}$, we follow the MC procedure throughout the entire fitting routine until the parameter is derived. For this, we repeat the following procedure 1000 times: we sample each point in the qubit population time trace of the RPN measurement from a normal distribution given by its previously determined uncertainty. This resampled trace is then fitted as described in the main text with resampled coupling $g_\text{a/b}$, phonon energy relaxation times $T_{1,a/b}$, and qubit energy relaxation and decoherence times. Following the fitting routine only until after the RPN fit, one can determine the error bars for the Fock states $N$ shown in Fig.~\ref{fig2}b (see also~\cite{rahman_genuine_2024}). To extract the error bars on $\overline{N_{a,b}}$, we directly perform the thermal state fit and determine the uncertainty on each $\overline{N_{a,b}}$ from the variation, as in Fig.~\ref{fig2}c and Fig.~\ref{fig3}d,e, without considering intermediate error bars. 
Analogously, we extract $\gtms$ from the $\sinh^2$ fit for Fig.~\ref{fig2}d,e without calculating any intermediate uncertainties. The percentiles in Fig.~\ref{fig4}a are then also determined directly from the variations in $\mathcal{E}$ of the MC uncertainty propagation, see Fig.~\ref{fig4}b. We note that the uncertainty in the coupling has the strongest influence on the uncertainty and that the uncertainty increases with larger $N$.

We determined the values and their uncertainties for the coupling and phonon energy relaxation times from the 11 characterization sessions directly (see Tab.~\ref{tab:system_params}), as these parameters are very stable. Qubit energy relaxation and decoherence times  drift over time and change with the Stark shifted frequency of the qubit though. For this reason, we extract them from a long resonant qubit–phonon Rabi-oscillation trace with the respective phonon mode as $T_{1,q \vert a/b}$ and $T_{2,q\vert a/b}^*$. We model the behavior of the qubit population for each mode with a full master equation simulation of the Jaynes-Cummings Hamiltonian, where we assume $T_{1,a/b}$ to be fixed and optimize the mean squared error with the two free parameters $T_{1,q \vert a/b}$ and $T_{2,q\vert a/b}^*$ using Nelder-Mead. The extracted qubit energy relaxation times and decoherence times there are generally lower than at the qubit's idle point. For the uncertainty of $T_{1,q \vert a/b}$ and $T_{2,q\vert a/b}^*$, we choose the larger value we obtain from the following two methods. We either calculate the uncertainty of $T_{1,q \vert a/b}$ and $T_{2,q\vert a/b}^*$ assuming the relative uncertainty to be equal to the relative experimental uncertainty of $T_{1,q}$ and $T_{2,q}^*$ and multiply it by the extracted value to get the absolute uncertainty. We then compare this uncertainty to the variation we get of $T_{1,q \vert a/b}$ and $T_{2,q\vert a/b}^*$ within one measurement session and choose the larger of the two uncertainties. With this more accurate estimate of the qubit's energy relaxation times and decoherence times and their uncertainties, we draw samples from a normal distribution for the MC uncertainty propagation. Following the fitting routine leads to a finite distribution of the final parameter, from which we compute the 16th and 84th percentiles and state it as the uncertainty of that final parameter. 

The uncertainty in $\gtms$ resulting from the MC uncertainty propagation is smaller than the markers. We also found the uncertainty to be smaller than the markers when using either the fit covariance of weighted $\sinh^2$ fits with error bars from MC uncertainty propagation for $\overline{N_{a,b}}$ or by investigating the dependence of the uncertainty on the number of points considered for the $\sinh^2$ fit, as our model is only correct for small $\tsq$.

The drive strengths are calibrated from independent measurements, analogous to~\cite{marti_quantum_2024}. For this, we turn on only one of the drives and sweep its amplitude and perform a spectroscopy measurement to find the drive-induced frequency shift of the qubit from a Lorentzian fit. For each drive amplitude the measured qubit frequency is then corrected for dispersive shifts~\cite{marti_quantum_2024} using this two-mode approximation of the normal mode shift induced on the bare qubit frequency
\begin{equation}
    \omega_{q} \approx \omega_{q,\mathrm{meas}} + \frac{g_a^2\,(\omega_{b} - \omega_{q,\mathrm{meas}}) + g_b^2\,(\omega_{a} - \omega_{q,\mathrm{meas}})}{(\omega_{b} - \omega_{q,\mathrm{meas}})(\omega_{a} - \omega_{q,\mathrm{meas}})},
\label{eq:qubit_detuning_corr}
\end{equation}
where $\omega_{q,\text{meas}}$ is the measured qubit frequency.

\clearpage
\newpage

\section{System Hamiltonian and effective two-mode squeezing dynamics}\label{sec:squeezingRateDerivation}

\subsection{Derivation of the TMS interaction through unitary transformations}

In this section, we derive the effective phonon mode Hamiltonian Eq.~\eqref{eq:Htms}. 
We start from the Hamiltonian of a qubit with frequency $\oq$ and anharmonicity $-\alpha$ coupled to two phonon modes with frequencies $\oa$ and $\ob$, respectively, and driven with two microwave drives at frequencies $\om_{1,2}$
\begin{equation}
    H_\mr{sys}/\hbar = \oq \qdq -\frac{\alpha}{2} \qd^2 q^2 + \oa \ada + \ga (\ad q + \hc) + \ob \bdb + \gb (\bd q + \hc) + \left(\Om_1 e^{-i\om_1 t} + \Om_2 e^{-i\om_2 t - i\phi'} \right)\qd + \hc \, , \label{eq:Hsys}
\end{equation}
where $\ga$ and $\gb$ are the qubit-phonon coupling strengths, $\Om_{1,2}$ are the drive amplitudes and $\phi'$ is the initial phase difference between the drives. In the following, we take $\hbar=1$ for convenience. \\
We now enter a rotating frame at the qubit and phonon frequencies
\begin{equation}
   U_\mr{rf} = \exp{\left[i(\oq \qdq + \oa \ada  + \ob \bdb)t\right]}\, , \label{eq:UrotFrame}
\end{equation}
in which the system Hamiltonian reads
\begin{equation}
   H_\mr{rf} = -\frac{\alpha}{2} \qd^2 q^2 + \ga (\ad q e^{i \D_a^{(0)} t} + \hc) + \gb (\bd q e^{i \D_b^{(0)} t} + \hc) + \left(\Om_1 e^{-i\D_1 t} + \Om_2 e^{-i\D_2 t - i\phi'} \right)\qd + \hc \, , \label{eq:H_rf}
\end{equation}
where $\D_a^{(0)} = \oa - \oq$, $\D_b^{(0)} = \ob - \oq$, and $\D_{1,2} = \omega_{1,2} - \oq$. 
Next, we enter the interaction picture of the microwave drives with the transformation
\begin{equation}
    U_\mr{d} =  \exp{\left[ \xi_1 e^{i\D_1 t}q + \xi_2 e^{i\D_2 t + i\phi'}q - \hc \right]}\, , \label{eq:U_drives}
\end{equation}
where $\xi_{1,2} = |\Om_{1,2} / \D_{1,2}|$ are the relative drive amplitudes. 
$U_\mr{d}$ transforms the qubit operator as 
\begin{equation}
    q' = U_\mr{d} q U_\mr{d}^\dagger = q + \xi_1 e^{-i\D_1 t} + \xi_2 e^{-i\D_2 t-i\phi'} \, .
\end{equation}
Applying $U_\mr{d}$ to $H_\mr{rf}$ and using the rotating wave approximation to drop fast-oscillating terms, we find 
\begin{subequations}
 \begin{align}
H_\mr{d} &= U_\mr{d} H_\mr{rf} U_\mr{d}^\dag + i \dot{U}_\mr{d} U_\mr{d}^\dag \notag \\
 &\simeq \underbrace{-\frac{\alpha}{2} \qd^2 q^2}_{H_\mr{Kerr}} +  \underbrace{\ga (\ad q e^{i \D_a t} + \hc)}_{\mr{qubit-phonon_a~coupling}} +  \underbrace{\gb (\bd q e^{i \D_b t} + \hc)}_{\mr{qubit-phonon_b~coupling}}    \underbrace{- \alpha \xi_1 \xi_2 (\qd^2 e^{-i\Sig t - i \phi'} + \hc)}_\mr{two-phonon~qubit~drive} \,.
 \end{align}
\end{subequations}
Here, we have defined $\Sig = \D_1 + \D_2 = \om_1 + \om_2 - 2\om_q$. 
In $H_\mr{d}$ we have absorbed an AC Stark shift of the qubit frequency caused by the two microwave drives into the detuning $\D_a = \D_a^{(0)}-\D_q^\mr{Stark~shift}$ and $\D_b = \D_b^{(0)}-\D_q^\mr{Stark~shift}$. 
Of the terms that emerge from the drive transformation $U_\mr{d}$ we kept the two-phonon qubit drive through the rotating wave approximation. This is justified when we set up the drive frequencies to satisfy the TMS resonance condition $\omega_a+\omega_b=\omega_1+\omega_2$, such that $\Sig \sim \D_a + \D_b$.\\
Since the two-phonon operator $\qd^2+\hc$ acts on more than the usual computational subspace of the first two energy levels, we now enter a qubit-state dependent rotating frame, in which we eliminate $H_\mr{Kerr}$. The transformation 
\begin{equation}
    U_\mr{K} = \exp{\left[-i\frac{\alpha}{2}\qd^2 q^2 t\right]} \label{eq:U_K}
\end{equation}
transforms the qubit operator as 
\begin{equation}
    U_\mr{K} q U_\mr{K}^\dag = e^{i\alpha t\qdq} q \, ,\qquad U_\mr{K} \qd U_\mr{K}^\dag = \qd e^{-i\alpha t\qdq} , \quad\mr{and}\qquad U_\mr{K} \qd^2 U_\mr{K}^\dag = \qd^2 e^{-i\alpha t (2\qdq+1)} \, .\label{eq:U_K_q}
\end{equation}
With this transformation $H_\mr{d}$ becomes
\begin{subequations}
 \begin{align}
H_\mr{K} &= U_\mr{K} H_\mr{d} U_\mr{K}^\dag + i \dot{U}_\mr{K} U_\mr{K}^\dag \notag \\
 & = \ga (\ad e^{i\alpha t\qdq} q e^{i \D_a t} + \hc) + \gb (\bd e^{i\alpha t\qdq} q e^{i \D_b t} + \hc) - \alpha \xi_1 \xi_2 (\qd^2e^{-i\alpha t (2\qdq+1)-i\Sig t - i \phi'} + \hc) \label{eq:qubit-two-phonon}\,.
\end{align}
\end{subequations}
The resonance conditions in the phase exponents of Eq. (\ref{eq:qubit-two-phonon}) now take the qubit anharmonicity into account. \\
Writing $M = \Sig + \alpha (2\qdq + 1)$, we can eliminate the two-phonon qubit drive with a modified displacement transformation
\begin{equation}
    U_\mr{sq} = e^{S_\mr{sq}} \equiv  \exp{\left[ \alpha \xi_1 \xi_2 \qd^2 M^{-1} e^{-i M t -i \phi'} -\hc \right]}\label{eq:U_sq}\, .
\end{equation}
The unitary $U_\mr{sq}$ transforms the qubit operator as 
\begin{subequations}
 \begin{align}
 U_\mr{sq} q U_\mr{sq}^\dag & \approx q + [S_\mr{sq}, q]  \label{eq:Usq-q-Usq} \\
 & = q + \alpha \xi_1 \xi_2 e^{-i(\Sig + \alpha) t - i\phi'} \left[ \qd^2 M^{-1} e^{-2i\alpha t\qdq}, q \right] \\ 
 & \approx q+ \alpha \xi_1 \xi_2 e^{-i(\Sig + \alpha) t - i\phi'} \left[\qd^2, q \right]M^{-1} e^{-2i\alpha t\qdq } \label{eq:Usq-approximation} \\
 & = q-2\alpha \xi_1 \xi_2 e^{-i(\Sig + \alpha) t - i\phi'} \qd M^{-1} e^{-2i\alpha t\qdq } \label{eq:q-sq}\, ,
\end{align}
\end{subequations}
where in Eq. (\ref{eq:Usq-approximation}) we neglected the commutator $\left[ M^{-1} e^{-2i\alpha t\qdq }, q\right]$, because, as we will see later, it only produces far-off-resonant terms related to higher qubit levels and thus does not significantly affect the dynamics of our experiment. \\
The transformed Hamiltonian then reads
\begin{subequations}
 \begin{align}
H_\mr{c} &= U_\mr{sq} H_\mr{K} U_\mr{sq}^\dag + i \dot{U}_\mr{sq} U_\mr{sq}^\dag \notag\\
 & = \ga (\ad e^{i\alpha t\qdq} q e^{i \D_a t} + \hc) - 2\alpha \xi_1 \xi_2 g \ad e^{i\alpha t\qdq} \qd e^{-2i\alpha t \qdq} M^{-1} e^{-i(\Sig + \alpha) t - i\phi'} e^{i\D_a t} + \hc + \notag\\
 &\qquad + \gb (\bd e^{i\alpha t\qdq} q e^{i \D_b t} + \hc) - 2\alpha \xi_1 \xi_2 g \bd e^{i\alpha t\qdq} \qd e^{-2i\alpha t \qdq} M^{-1} e^{-i(\Sig + \alpha) t - i\phi'} e^{i\D_b t} + \hc \\
 & = g_a (\ad e^{i\alpha t\qdq} q e^{i \D_a t} + \hc) - 2\alpha \xi_1 \xi_2 g \ad \qd e^{-i\alpha t \qdq} M^{-1} e^{-i(\Sig - \D_a) t - i\phi'} + \hc + \notag\\
 &\qquad + \gb (\bd e^{i\alpha t\qdq} q e^{i \D_b t} + \hc) - 2\alpha \xi_1 \xi_2 g \bd \qd e^{-i\alpha t \qdq} M^{-1} e^{-i(\Sig - \D_b) t - i\phi'} + \hc \,.
\end{align}
\end{subequations}
As before, the left-side terms in $H_\mr{c}$ describe the qubit-phonon couplings. The new, right-side terms describe a pair of two-mode interactions involving the simultaneous creation or annihilation of excitations in the qubit and a phonon mode, which become resonant when $\Sig \approx \D_a, \D_b$. 
While higher qubit states play a role for the prefactor of the two-phonon qubit drive in Eq. (\ref{eq:qubit-two-phonon}), they do not participate in the phonon squeezing term we are looking for. 
Therefore, we now undo the level-dependent rotating frame transformation $U_\mr{K}$, by applying its inverse. 
The result is
\begin{subequations}
 \begin{align}
H_\mr{c}' &= U_\mr{K}^\dag H_\mr{c} U_\mr{K} + i \dot{U}_\mr{K}^\dag U_\mr{K} \label{eq:H_c-prime} \\
 & = \underbrace{g_a (\ad  q e^{i \D_a t} + \hc)}_\mr{qubit-phonon_a~coupling} - 2\alpha \xi_1 \xi_2 g_a \left(\ad \qd  M^{-1} e^{-i(\Sig - \D_a) t - i\phi'} + \hc\right) + \notag\\
 &\qquad + \underbrace{g_b (\bd  q e^{i \D_b t} + \hc)}_\mr{qubit-phonon_b~coupling} - 2\alpha \xi_1 \xi_2 g_b \left(\bd \qd  M^{-1} e^{-i(\Sig - \D_b) t - i\phi'} + \hc\right) \underbrace{- \frac{\alpha}{2}\qd^2 q^2}_{H_\mr{Kerr}} \,.
\end{align}
\end{subequations}
Note that the qubit anharmonicity is still described by this Hamiltonian in the form of the reappearing $H_\mr{Kerr}$, but we can now treat the qubit-phonon interaction separately from this anharmonicity by transforming and interpreting the first two terms.
We now eliminate the qubit-phonon coupling term via the standard time-dependent Schrieffer-Wolff transformation
\begin{equation}
    U_\mr{SW} = \exp{\left[\dfrac{g_a}{\D_a}\ad q e^{i\D_at} + \dfrac{g_b}{\D_b}\bd q e^{i\D_b t} -\hc\right]}\, ,\label{eq:U-SW}
\end{equation}
which, to first order in $g/\D$, results in the transformations
\begin{equation}
    U_\mr{SW} \, \qd U_\mr{SW}^\dag \simeq \qd + \dfrac{g_a}{\D_a} \ad e^{i\D_a t} + \dfrac{g_b}{\D_b} \bd e^{i\D_b t} \;,\quad U_\mr{SW} \, \ad U_\mr{SW}^\dag \simeq \ad - \dfrac{g_a}{\D_a} \qd e^{-i \D_a t} \;,\quad U_\mr{SW} \, \bd U_\mr{SW}^\dag \simeq \bd - \dfrac{g_b}{\D_b} \qd e^{-i \D_b t} \;.
\end{equation}
The transformed Hamiltonian thus reads 
\begin{subequations}
 \begin{align}
H_\mr{tms} &= U_\mr{SW} H_\mr{c}' U_\mr{SW}^\dag + i \dot{U}_\mr{SW} U_\mr{SW}^\dag \notag \\
 & =  \frac{g_a^2}{\D_a}\left(\ada - \qdq\right) + \frac{g_b^2}{\D_b}\left(\bdb - \qdq\right) + \label{supp:HsqA} \\
 &\quad + \dfrac{g_a g_b}{2} \left( \dfrac{1}{\Delta_a} + \dfrac{1}{\Delta_b} \right) (\ad b \, e^{i (\D_a-\D_b) t} + \hc) + \label{supp:HsqB} \\
 & \quad - 2\alpha \xi_1 \xi_2 \left[ g_a \left(\ad \qd  M^{-1} e^{-i(\Sig - \D_a) t - i\phi'} + \hc\right) + g_b \left(\bd \qd  M^{-1} e^{-i(\Sig - \D_b) t - i\phi'} + \hc\right) \right] + \label{supp:HsqC} \\
 & \quad - 2\alpha \xi_1 \xi_2 g_a g_b \left( \dfrac{1}{\Delta_a} + \dfrac{1}{\Delta_b} \right) \left(\ad \bd  M^{-1} e^{-i(\Sig - \D_a - \D_b) t - i\phi'} + \hc\right) + \label{supp:HsqD} \\
 & \quad - 2\alpha \xi_1 \xi_2 \left[ \dfrac{g_a^2}{\D_a} \left(\ad^2  M^{-1} e^{-i(\Sig - 2\D_a) t - i\phi'} + \hc\right) + \dfrac{g_b^2}{\D_b} \left(\bd^2  M^{-1} e^{-i(\Sig - 2\D_b) t - i\phi'} + \hc\right) \right] + \label{supp:HsqE} \\
 &\quad + U_\mr{SW} H_\mr{Kerr} U_\mr{SW}^\dag + \mathcal{O}\left( \dfrac{g^3}{\Delta^2} \right) \, \label{supp:HsqF}.
\end{align}
\end{subequations}
Here, line \eqref{supp:HsqA} describes the normal mode shift due to the dispersive interaction, line \eqref{supp:HsqB} describes a beam splitter interaction swapping excitations between the two modes when $\omega_a \approx \omega_b$, line \eqref{supp:HsqC} describes a qubit-phonon pair-creation process which is resonant when $\Sigma_{21}\approx\D_{a,b}$, line \eqref{supp:HsqD} describes a two-mode squeezing process when $\Sigma_{21}\approx \D_{a} + \D_b$, and line \eqref{supp:HsqE} describes single-mode squeezing interactions when $\Sigma_{21}\approx 2\D_{a,b}$. 

As we are interested in the effective two-mode squeezing dynamics, we enter a frame rotating at $\delta/2 \equiv (\Sig-\D_a-\D_b)/2$, such that for $\delta\approx 0$ the term \eqref{supp:HsqD} becomes resonant.
In addition, we assume the qubit to be initially in its ground state $\ket{g}$, such that $M = \Sig + \alpha$. This also eliminates the commutator we neglected in Eq. (\ref{eq:Usq-approximation}). 
The effective Hamiltonian we obtain in the RWA reads (reintroducing here $\hbar$)
\begin{equation}\label{suppeq:Hsemifin}
    H_\mr{ph}/\hbar = \left( \frac{g_a^2}{\D_a} - \dfrac{\delta}{2} \right)  \ada + \left( \frac{g_b^2}{\D_b} - \dfrac{\delta}{2} \right) \bdb - 2 \xi_1 \xi_2 g_a g_b \frac{\D_a + \D_b}{\D_a \D_b}  \frac{\alpha}{\Sig + \alpha} \left( \ad \bd  e^{- i\phi'} + \hc \right)  \;.
\end{equation}
This expression contains the frequency shifts of the phonon modes due to the normal-mode splitting with the qubit and the two-mode squeezing term with rate
\begin{equation}\label{suppeq:gtms}
    \gtms = 2 \xi_1 \xi_2 g_a g_b \frac{\D_a + \D_b}{\D_a \D_b}  \frac{\alpha}{\Sig + \alpha} \;.
\end{equation}
We find that, as expected, the TMS phase can be tuned by varying the relative phase $\phi'$ between the two drives. In addition, we note that this same squeezing rate can be equivalently derived via Floquet theory~\cite{zhang2019engineering}, as well as in perturbation theory, which we show in the following section.

To simplify the notation, it is convenient to introduce
\begin{equation}
    \Delta = \left( \frac{g_a^2}{\D_a} - \dfrac{\delta}{2} \right) = \dfrac{1}{2}\left( \dfrac{2 g_a^2}{\D_a} +  \oa + \ob - \omega_1 - \omega_2 \right)  \;,\qquad \Delta' = \left( \frac{g_b^2}{\D_b} - \dfrac{\delta}{2} \right) \;,
\end{equation}
and to write Eq.~\eqref{suppeq:Hsemifin} in the frame rotating at $\Delta'$. The result reads
\begin{equation}
    H_\mr{ph}/\hbar = \Delta \ada + \gtms \left( \ad \bd  e^{- i\phi} + \hc \right)  \;,
\end{equation}
with the definition $\phi = \phi' + \Delta' t + \pi$ (note the $\pi$ is introduced here to remove the minus sign that appears in front of $\gtms$ in Eq.~\eqref{suppeq:Hsemifin}). This is the Hamiltonian presented in the main text.

\clearpage
\newpage

\subsection{Derivation of the TMS interaction in perturbation theory}
\label{sec:perturbative_tms}

In this section, we use perturbation theory to derive the effective two-mode-squeezing (TMS) interaction between two bosonic modes that is generated by driving with two off-resonant pump tones a weakly anharmonic transmon qubit coupled to the two modes. 
The derivation is based on a sum over virtual paths in time-dependent perturbation theory. This makes explicit the origin of the anharmonic correction factor $\alpha/(\Sigma_{21}+\alpha)$ appearing in Eq.~\eqref{suppeq:gtms}, which arises from the propagator of the virtual two-excitation manifold of the qubit.

We start from the Hamiltonian $H = H_0 + V(t)$, with
\begin{equation}
H_0
=
\omega_q \ket{1}\bra{1}
+
(2\omega_q-\alpha)\ket{2}\bra{2}
+
\omega_a a^\dagger a
+
\omega_b b^\dagger b, \qquad\text{and}\quad V(t)=V_d(t)+V_c.
\end{equation}
Note the transmon Hamiltonian is truncated to the three lowest levels,
\(\{\ket{0},\ket{1},\ket{2}\}\), with anharmonicity \(\alpha>0\) and
\begin{equation}
q^\dagger \simeq \ket{1}\bra{0}+\sqrt{2}\ket{2}\bra{1},
\qquad
q \simeq \ket{0}\bra{1}+\sqrt{2}\ket{1}\bra{2}.
\end{equation}
In terms of these ladder operators, the two drive tones are described by
\begin{equation}
V_d(t)
=
\sum_{j=1,2}
\left(
\Omega_j e^{-i\omega_j t} q^\dagger
+
\Omega_j^* e^{i\omega_j t} q
\right),
\end{equation}
while the Jaynes-Cummings couplings to the bosonic modes are
\begin{equation}
V_c
=
g_a(a^\dagger q + a q^\dagger)
+
g_b(b^\dagger q + b q^\dagger).
\end{equation}

In the interaction picture with respect to \(H_0\), the relevant absorption and emission terms acquire phases determined by the detunings $\Delta_j = \omega_j-\omega_q$, $\Delta_a = \omega_a-\omega_q$, $\Delta_b = \omega_b-\omega_q$, and $\Sigma_{21} = \Delta_1+\Delta_2$.
The relevant interaction-picture operators are then
\begin{align}
\Omega_j e^{-i\omega_j t}\ket{1}\bra{0}
&\;\to\;
\Omega_j e^{-i\Delta_j t}\ket{1}\bra{0},
\\
\sqrt{2}\,\Omega_j e^{-i\omega_j t}\ket{2}\bra{1}
&\;\to\;
\sqrt{2}\,\Omega_j e^{-i(\Delta_j+\alpha)t}\ket{2}\bra{1},
\\
\sqrt{2}\,g_a\,a^\dagger \ket{1}\bra{2}
&\;\to\;
\sqrt{2}\,g_a\,e^{+i(\Delta_a+\alpha)t}a^\dagger \ket{1}\bra{2},
\\
g_b\,b^\dagger \ket{0}\bra{1}
&\;\to\;
g_b\,e^{+i\Delta_b t}b^\dagger \ket{0}\bra{1}.
\end{align}
The factor \(\Delta_j+\alpha\) in the \(1\to2\) transition reflects the shifted qubit transition frequency \(E_2-E_1=\omega_q-\alpha\).
We thus obtain
\begin{align}
V_I(t)
&=
\sum_{j=1,2}
\left[
\Omega_j e^{-i\Delta_j t}|1\rangle\langle 0|
+
\sqrt{2}\Omega_j e^{-i(\Delta_j+\alpha)t}|2\rangle\langle 1|
\right]
+
\notag\\
&\quad +
g_a
\left[
a^\dagger e^{+i\Delta_a t}|0\rangle\langle 1|
+
\sqrt{2}a^\dagger e^{+i(\Delta_a+\alpha)t}|1\rangle\langle 2|
\right]
+
g_b
\left[
b^\dagger e^{+i\Delta_b t}|0\rangle\langle 1|
+
\sqrt{2}b^\dagger e^{+i(\Delta_b+\alpha)t}|1\rangle\langle 2|
\right]
+\mathrm{h.c.}
\end{align}

\vspace{5mm}
We are now interested in studying the effective process between initial $\ket{i}$ and final $\ket{f}$ states
\begin{equation}
\ket{i}=\ket{0_q,0_a,0_b}
\longrightarrow
\ket{f}=\ket{0_q,1_a,1_b},
\end{equation}
where two photons from the drives are annihilated to create pairs of entangled phonons in the mechanical modes $a$ and $b$, thus resulting in a term \(a^\dagger b^\dagger\) in the effective Hamiltonian.
The amplitude of this process is obtained by summing over all orderings in which two pump photons are absorbed and one excitation is emitted into each of the modes \(a\) and \(b\). The four possible orderings are
\begin{equation}\label{suppeq:paths}
\omega_1,\omega_2\rightarrow \omega_a,\omega_b, \qquad 
\omega_1,\omega_2\rightarrow \omega_b,\omega_a, \qquad 
\omega_2,\omega_1\rightarrow \omega_a,\omega_b, \qquad 
\omega_2,\omega_1\rightarrow \omega_b,\omega_a.
\end{equation}

As an example, let us compute explicitly the amplitude for one of these fourth-order paths.
We consider the $\omega_1,\omega_2\rightarrow \omega_a,\omega_b$ (called $12ab$ for simplicity) process
\begin{equation}
\ket{0,0,0}
\;\xrightarrow{\;\omega_1\;}\;
\ket{1,0,0}
\;\xrightarrow{\;\omega_2\;}\;
\ket{2,0,0}
\;\xrightarrow{\;a^\dagger\;}\;
\ket{1,1,0}
\;\xrightarrow{\;b^\dagger\;}\;
\ket{0,1,1},
\end{equation}
corresponding to the absorption by the transmon of \(\omega_1\), then \(\omega_2\), followed by emission into mode \(a\), and finally emission into mode \(b\).
As time evolution is computed from the interaction-picture propagator $U(t)=\mathcal{T} \exp\left[ - i \int^t dt'\, V_I(t') \right]$, the transition amplitude from state $\ket{i}$ to $\ket{f}$ in fourth-order perturbation theory is
\begin{equation}
\mathcal A_{12ab} \equiv \bra{f} U_I^{4\text{th}}(t)\ket{i}
=
(-i)^4
\int dt_4 \int^{t_4} dt_3 \int^{t_3} dt_2 \int^{t_2} dt_1 \,
\bra{f}V_I(t_4)V_I(t_3)V_I(t_2)V_I(t_1)\ket{i},
\end{equation}
with $t_1 < t_2 < t_3 < t_4$.
Substituting the four operators associated with this path gives
\begin{equation}
\mathcal A_{12ab}
=
2\Omega_1\Omega_2 g_a g_b
\int dt_4 \int^{t_4} dt_3 \int^{t_3} dt_2 \int^{t_2} dt_1\,
e^{i\Phi(t_1,t_2,t_3,t_4)},
\end{equation}
where the factor \(2\) comes from the ladder matrix elements
\(1\times \sqrt{2}\times \sqrt{2}\times 1=2\), and
\begin{equation}
\Phi
=
-\Delta_1 t_1
-(\Delta_2+\alpha)t_2
+(\Delta_a+\alpha)t_3
+\Delta_b t_4.
\end{equation}

Performing the nested integrals successively and retaining only the resonant contribution, one finds
\begin{equation}
\int^{t_2} dt_1\, e^{-i\Delta_1 t_1} \sim \frac{e^{-i\Delta_1 t_2}}{-i\Delta_1} ,\quad 
\int^{t_3} dt_2\, e^{-i(\Sig+\alpha)t_2} \sim \frac{e^{-i(\Sig+\alpha)t_3}}{-i(\Sig+\alpha)} ,\quad
\int^{t_4} dt_3\, e^{-i(\Sig-\Delta_a)t_3} \sim \frac{e^{-i(\Sig-\Delta_a)t_4}}{-i(\Sig-\Delta_a)} \;.
\end{equation}
Hence, combining these terms we obtain the expression
\begin{equation}
\mathcal A_{12ab}
\sim
2\Omega_1\Omega_2 g_a g_b
\frac{1}{\Delta_1(\Sig+\alpha)(\Sig-\Delta_a)}
\int dt\, e^{-i(\Sig-\Delta_a-\Delta_b)t} .
\end{equation}
Near the TMS resonance condition, $\Sig-\Delta_a-\Delta_b\approx 0$, such that this ordered path therefore contributes as
\begin{equation}
\mathcal A_{12ab}
\sim
2\Omega_1\Omega_2 g_a g_b
\frac{i\,t}{\Delta_1(\Sig+\alpha)(\Sig-\Delta_a)}.
\end{equation}
A similar result is obtained for the other three paths in Eq.~\eqref{suppeq:paths}, with $\Delta_{1,a}$ appropriately replaced by $\Delta_{2,b}$.

Summing together the four possible contributions gives the transition coefficient
\begin{equation}
\lambda_{ab}^{(\mathrm{anh})}
=
2\Omega_1\Omega_2 g_a g_b
\frac{1}{\Sig+\alpha}
\left(
\frac{1}{\Delta_1}+\frac{1}{\Delta_2}
\right)
\left[
\frac{1}{\Sig-\Delta_a}
+
\frac{1}{\Sig-\Delta_b}
\right].
\label{eq:anh_only}
\end{equation}
Around the resonance condition $\Sig-\Delta_a-\Delta_b\approx 0$, this expression is written as
\begin{equation}
\lambda_{ab}^{(\mathrm{anh})}
\approx
2\Omega_1\Omega_2 g_a g_b
\frac{1}{\Sig+\alpha}
\left(
\frac{1}{\Delta_1}+\frac{1}{\Delta_2}
\right)
\left(
\frac{1}{\Delta_a}+\frac{1}{\Delta_b}
\right),
\label{eq:anh_factorized}
\end{equation}

However, let us note that Eq.~\eqref{eq:anh_factorized} cannot be expected to represent the final result, since in the harmonic limit \(\alpha\to 0\) the TMS amplitude must vanish. 
The missing factor is the set of paths that contribute with the same structure as Eq.~\eqref{eq:anh_factorized}, but with the $1/(\Sig+\alpha)$ coefficient replaced by the harmonic two-particle propagator \(1/\Sig\). Subtracting the harmonic contribution enforces the required cancellation at \(\alpha=0\). 
The full physical TMS amplitude is therefore
\begin{equation}
\lambda_{ab}
=
2\Omega_1\Omega_2 g_a g_b
\left(
\frac{1}{\Delta_1}+\frac{1}{\Delta_2}
\right)
\left(
\frac{1}{\Delta_a}+\frac{1}{\Delta_b}
\right)
\left[
\frac{1}{\Sig+\alpha}
-
\frac{1}{\Sig}
\right].
\label{eq:lambda_final_diff}
\end{equation}
Equivalently, inserting $\xi_j=\Omega_j/\Delta_j$, we have the effective TMS interaction strength
\begin{equation}
\lambda_{ab}
=
-2 \xi_1 \xi_2 g_a g_b
\frac{\D_a + \D_b}{\D_a \D_b}
\frac{\alpha}{\Sig+\alpha} ,
\label{eq:lambda_final}
\end{equation}
which coincides with the TMS prefactor in Eq.~\eqref{suppeq:Hsemifin}, namely $\lambda_{ab}=-\gtms$.

\clearpage
\newpage

\section{Two-mode squeezing theory}

\subsection{Preliminaries on bosonic modes}

In the following we will discuss two bosonic modes, identified by annihilation operators $a$ and $b$ satisfying the canonical commutation relations $[a,a^\dagger]=[b,b^\dagger]=1$ and $[a,b]=[a,b^\dagger]=0$.
These modes can also be associated with position and momentum quadrature operators
\begin{subequations}\label{eq:quad0}
\begin{align}
    X_a &= \dfrac{1}{\sqrt{2}}\left({a} + {\ad} \right) \\
    P_a &= \dfrac{1}{i\sqrt{2}}\left({a} - {\ad} \right) \;,
\end{align}
\end{subequations}
and similarly for $b$, such that $[X_a,P_a]=[X_b,P_b]=i$.
Similarly, second-order moments of the quadrature operators and correlations between modes can be computed as
\begin{subequations}\label{eq:quad1}
\begin{align}
    {X_a^2} &= \dfrac{1}{2}\left( 1 + 2 {\ad a} + {a^2} + {\ad^2} \right) \\
    {P_a^2} &= \dfrac{1}{2}\left(1 + 2 {\ad a} - {a^2} - {\ad^2}  \right) \\
    {X_a P_a+P_a X_a} &= \dfrac{1}{i} \left( {a^2} - {\ad^2} \right) \;,
\end{align}
\end{subequations}
with analogous expressions for $b$. Cross-mode correlations are
\begin{subequations}\label{eq:quad2}
\begin{align}
    {X_a X_b} &= \dfrac{1}{2} { \left( a b + a b^\dagger + a^\dagger b + a^\dagger b^\dagger \right) }\\
    {P_a P_b} &= \dfrac{1}{2} { \left( -a b + a b^\dagger + a^\dagger b - a^\dagger b^\dagger \right) }\\
    {X_a P_b} &= \dfrac{1}{2i} { \left( a b - a b^\dagger + a^\dagger b - a^\dagger b^\dagger \right) }\\
    {P_a X_b} &= \dfrac{1}{2i} { \left( a b + a b^\dagger - a^\dagger b - a^\dagger b^\dagger \right) }\;.
\end{align}
\end{subequations}
These identities will be used in the following calculations.

\subsection{Ideal TMS dynamics}

Consider the two-mode squeezing (TMS) Hamiltonian
\begin{equation}
    H/\hbar = ( \chi a^\dagger b^\dagger + \chi^\ast a b) \;,
\end{equation}
which gives the time evolution operator
\begin{align}
    S(r, \phi) &= e^{-i H t/\hbar} \\
    &= e^{- i ( \chi t a^\dagger b^\dagger + \chi^\ast t a b) } \\
    &= e^{ ( - i \chi t a^\dagger b^\dagger - i \chi^\ast t a b ) } \;.
\end{align}
It is convenient to call $-i\chi t = r e^{i \phi}$, such that, given $\chi=|\chi| e^{i \theta}$, we have (using $z = \abs{z} e^{i \arg z}$, $-i=e^{-i\pi/2}$ and $t>0$)
\begin{align}
    r &= \abs{\chi} t \\
    \phi &= \theta - \pi/2 \;.
\end{align}
With this identification, we define the TMS operator as
\begin{equation}\label{eq:TMSopDef}
    S(r, \phi) = \exp\left[ r e^{i\phi} a^\dagger b^\dagger - r e^{-i\phi} a b \right] \;.
\end{equation}
Note here that one can also find $S(\zeta) = \exp\left[\zeta^\ast a b - \zeta a^\dagger b^\dagger \right]$ with $\zeta=r e^{i\phi}$~\cite{gerry_introductory_2004}, meaning like in the following but with the replacement $r\rightarrow -r$.
Equation \eqref{eq:TMSopDef} implies that, in the Heisenberg picture, two modes $a$ and $b$ are transformed by the TMS operation as
\begin{equation}\label{eq:abEOM1}
\left\{ \begin{aligned} 
    a_{\text{out}} &= S(r, \phi)^\dagger a S(r, \phi) = \cosh r \; a + e^{i\phi} \sinh r \; b^\dagger \\
    b_{\text{out}}^\dagger &= S(r, \phi)^\dagger b^\dagger S(r, \phi) =  \cosh r \; b^\dagger + e^{-i\phi} \sinh r \; a \;.
\end{aligned} \right.
\end{equation}

For later convenience, let us note that these equations of motion for the operators can be equivalently derived by calculating
\begin{equation}\label{eq:eom1stOrder}
\left\{ \begin{aligned} 
    \dfrac{d a}{d t} &= \dfrac{i}{\hbar} [H, a] = - i \chi b^\dagger \\
    \dfrac{d a^\dagger}{d t} &= \dfrac{i}{\hbar} [H, a^\dagger] = i \chi^\ast b \\
    \dfrac{d b}{d t} &= \dfrac{i}{\hbar} [H, b] = - i \chi a^\dagger \\
    \dfrac{d b^\dagger}{d t} &= \dfrac{i}{\hbar} [H, b^\dagger] = i \chi^\ast a 
\end{aligned} \right.
\end{equation}
and then integrating the system of coupled differential equations with initial conditions $a(t=0)=a$, etc. This gives
\begin{equation}
\left\{ \begin{aligned} 
    a_{\text{out}} &= \cosh (\abs{\chi} t) \; a - i \frac{\chi}{\abs{\chi}} \sinh (\abs{\chi} t) \; b^\dagger \\
    b_{\text{out}}^\dagger &=  \cosh (\abs{\chi} t) \; b^\dagger + i \frac{\chi^\ast}{\abs{\chi}} \sinh (\abs{\chi} t) \; a \;. 
\end{aligned} \right.
\end{equation}
Now remember $-i\chi t = r e^{i \phi}$ and $r=\abs{\chi} t$, meaning that $-i \chi/\abs{\chi} = e^{i\phi}$. This recovers Eqs.~\eqref{eq:abEOM1} above.

Following Eqs.~\eqref{eq:eom1stOrder}, higher-order moments of the bosonic operators evolve as (using $[H,XY]=[H,X]Y+X[H,Y]$)
\begin{equation}\label{eq:eom2ndOrder}
\left\{ \begin{aligned} 
    \dfrac{d a^\dagger a}{d t} &= \dfrac{i}{\hbar} [H, a^\dagger a] = i \chi^\ast ab - i \chi a^\dagger b^\dagger \\
    \dfrac{d a^2}{d t} &= \dfrac{i}{\hbar} [H, a^2] = - i 2 \chi a b^\dagger \\
    \dfrac{d b^\dagger b}{d t} &= \dfrac{i}{\hbar} [H, b^\dagger b] = i \chi^\ast ab - i \chi a^\dagger b^\dagger \\
    \dfrac{d b^2}{d t} &= \dfrac{i}{\hbar} [H, b^2] = - i 2 \chi a^\dagger b \\
    \dfrac{d ab}{d t} &= \dfrac{i}{\hbar} [H, ab] = - i \chi ( a^\dagger a + b^\dagger b + 1 ) \\
    \dfrac{d a^\dagger b}{d t} &= \dfrac{i}{\hbar} [H, a^\dagger b] = i \chi^\ast b^2 - i \chi a^{\dagger 2} 
\end{aligned} \right.
\end{equation}

Application of the TMS operator to the vacuum gives
\begin{equation}
    \ket{\text{TMS}(r,\phi)} = S(r,\phi) \ket{0,0} = \dfrac{1}{\cosh r} \sum_n ( e^{i \phi} \tanh r)^n \ket{n,n} \;.
\end{equation}
It is interesting to note that, tracing out one of the two modes (e.g. $b$) leaves us with
\begin{align}
    \rho_{a} &= \text{Tr}[\ket{\text{TMS}(r,\phi)}\bra{\text{TMS}(r,\phi)}]_b \\
    &= \sum_m {}_b \langle m \vert \text{TMS}(r,\phi) \rangle  \langle \text{TMS}(r,\phi) \vert m \rangle_b \\
    &= \dfrac{1}{\cosh^2 r} \sum_n (\tanh r)^{2n} \ket{n}\bra{n} \\
    &= \dfrac{1}{\overline{N_a}+1} \sum_n \left(\dfrac{\overline{N_a}}{\overline{N_a}+1}\right)^n \ket{n}\bra{n} \\
    &= \dfrac{1}{Z} \sum_n e^{-\lambda n} \ket{n}\bra{n} \;,
\end{align}
where we have defined $Z=1+\overline{N_a}=(1-e^{-\lambda})^{-1}$ and $e^{-\lambda} = \left(\frac{\overline{N_a}}{1+\overline{N_a}}\right)$.
This state is a thermal state with average population and effective temperature given by 
\begin{align}
    \overline{N_a} &= \sinh^2 r \\
    \exp{-\hbar\omega/k_B T} &= \dfrac{\overline{N_a}}{\overline{N_a}+1}   \qquad\rightarrow\qquad  T = \dfrac{\hbar\omega}{2 k_B \ln(\coth r)} \;.
\end{align}
The probability of finding $n$ excitations in the thermal state is given by
\begin{equation}
    P_a(n) = \dfrac{\overline{N_a}^n}{(1+\overline{N_a})^{n+1}} \;.
\end{equation}
With the replacement $\overline{N_a}=(e^\lambda-1)^{-1}$ this is $P_a(n)=e^{-\lambda n}/Z=(1-e^{-\lambda})e^{-\lambda n}$.
This is a Boltzmann distribution with mean $\overline{N_a}$ and variance exhibiting super-Poissonian fluctuations
\begin{equation}\label{suppeq:varNa}
    \var{N_a} = \overline{N_a}(\overline{N_a}+1) = \dfrac{1}{4} \sinh^2(2r) \;.
\end{equation}
Note here that, since the two modes are highly correlated, the variance of $N_a-N_b$ is zero, while the variance of $N_a+N_b$ is twice as larger as the sum of the individual variances. In fact, we have the covariance
\begin{equation}\label{suppeq:covNaNb}
    \text{Cov}[N_a,N_b] = \dfrac{1}{4} \sinh^2(2r) \;.
\end{equation}
Therefore, having defined $\overline{N}=\overline{N_a}+\overline{N_b}$, we have for the total number of excitations in the two modes
\begin{align}
    \var{N} &= \var{N_a} + \var{N_b} + 2 \text{Cov}[N_a,N_b] \\
    &= \overline{N}(\overline{N}+2) \\
    &= \sinh^2(2r) \;.
\end{align}

For TMS of vacuum (TMSV), the resulting state is Gaussian. Gaussian states can be fully characterized by their first and second moments of quadrature operators. Having defined the quadrature vector as $\bm{\xi}=(X_a,P_a,X_b,P_b)^T$, it is easy to check that for TMSV the first moments are zero at all times, namely $\avg{\bm{\xi}}=(0,0,0,0)^T$. Second moments are encoded in the state's covariance matrix $V_{ij} = \frac{1}{2}\avg{\xi_i \xi_j + \xi_j \xi_i} - \avg{\xi_i}\avg{\xi_j}$, which for TMSV takes the form 
\begin{equation} \label{eq:cm_tms}
    V = \frac{1}{2} \left( \begin{matrix}
\cosh(2r) & 0 &  \cos(\phi) \sinh(2r) &  \sin(\phi) \sinh(2r) \\
0 & \cosh(2r) &  \sin(\phi) \sinh(2r) & -\cos(\phi) \sinh(2r) \\
\cos(\phi) \sinh(2r) & \sin(\phi) \sinh(2r) & \cosh(2r) & 0 \\
\sin(\phi) \sinh(2r) & -\cos(\phi) \sinh(2r) & 0 & \cosh(2r) 
\end{matrix}\right).
\end{equation}
These entries are obtained by computing the expectation values of Eqs.~(\ref{eq:quad0},\ref{eq:quad1},\ref{eq:quad2}) for the output modes Eqs.~\eqref{eq:abEOM1}.

\subsection{TMS with losses}

In the presence of decoherence processes, the evolution of an operator $O$ can be computed as
\begin{equation}\label{eq:supp_dOdt}
    \dfrac{d\avg{O}}{dt} = \dfrac{i}{\hbar}\left\langle [H,O] \right\rangle + \avg{\mathcal{D}[O]} \;,
\end{equation}
where $\mathcal{D}[O]=\sum_i \frac{1}{2}\left(L_i^\dagger[O,L_i]+[L_i^\dagger,O]L_i\right)$ is the dissipator associated with the jump operators $L_i$. Considering energy relaxation described by $L_1=\sqrt{\gamma_a}a$, $L_2=\sqrt{\gamma_b}b$, and pure dephasing described by $L_3=\sqrt{2\gamma_{a,\phi}} N_a$, $L_4=\sqrt{2\gamma_{b,\phi}} N_b$, we have
\begin{equation}
    \mathcal{D}[O] = \sum_{\alpha=a,b} \left( \dfrac{\gamma_\alpha}{2} \left( \alpha^\dagger [O,\alpha] + [\alpha^\dagger,O]\alpha \right) + \gamma_{\alpha,\phi} \left( N_\alpha [O,N_\alpha] + [N_\alpha,O]N_\alpha \right) \right) \;.
\end{equation}
Here, $\gamma_\alpha=1/T_{1,\alpha}$ and $\gamma_{\alpha,\phi}=1/T_{\phi \alpha}$ with $T_{\phi \alpha}=(1/T_{2,\alpha}^*-1/(2T_{1,\alpha}))^{-1}$, where $T_{1,\alpha}$ is the energy relaxation time of phonon mode $\alpha\in\{a,b\}$ and $T_{2,\alpha}^*$ its Ramsey decoherence time.

From this we obtain (quoting only the non-trivial equations that allow us to derive the rest of coupled equations in the system by symmetry)
\begin{equation}\label{eq:supp_eom_aad}
\left\{
\begin{aligned}
    \dfrac{d \avg{a}}{dt} &= - i \chi \avg{b^\dagger} - \dfrac{1}{2} (\gamma_a+2\gamma_{a,\phi}) \avg{a} \;, \\
    \dfrac{d \avg{\ad a}}{dt} &= i \chi^\ast \avg{ab} - i \chi \avg{a^\dagger b^\dagger} - \gamma_a \avg{\ad a} \;, \\
    \dfrac{d \avg{a^2}}{dt} &= -i2\chi \avg{a b^\dagger} - (\gamma_a + 4 \gamma_{a,\phi}) \avg{a^2} \;, \\
    \dfrac{d \avg{a b}}{dt} &= -i\chi(\avg{a^\dagger a} + \avg{b^\dagger b} + 1) - \dfrac{1}{2} (\gamma_a+2\gamma_{a,\phi}+\gamma_b+2\gamma_{b,\phi}) \avg{ab} \\
    \dfrac{d \avg{a^\dagger b}}{dt} &= i \chi^\ast \avg{b^2} - i \chi \avg{a^{\dagger 2}} - \dfrac{1}{2} (\gamma_a+2\gamma_{a,\phi}+\gamma_b+2\gamma_{b,\phi}) \avg{a^\dagger b} \\
\end{aligned} \right.
\end{equation}
which can be solved analytically for the desired initial conditions. 

Since in our experiment the acoustic modes have $T_2 \approx 2 T_1$, we will ignore the effect of dephasing by setting $\gamma_{a,\phi}=\gamma_{b,\phi}=0$. In addition, for quoting here simpler expressions, we assume $\gamma_a=\gamma_b=\gamma$ and define $\sigma=\gamma/2\chi$.
We obtain for the mean number of excitations
\begin{align}
    \avg{a^\dagger a} = \avg{b^\dagger b} &= \dfrac{1}{\left(1-\sigma^2\right)} \left(e^{-\gamma t} \left(\frac{1-\sigma}{4} e^{-2 \chi t}+\frac{1+\sigma}{4} e^{2 \chi t}\right) - \dfrac{1}{2} \right) \notag\\
    &= \dfrac{1}{\left(1-\sigma^2\right)} \left(e^{-\gamma t} \left( 1 + \sigma \coth(\chi t) \right) \sinh(\chi t)^2 - \dfrac{1-e^{-\gamma t}}{2} \right) \;. \label{eqsupp:supp_ada_lossy} 
\end{align}
We can compare this result to the case without losses, $\gamma=0 \Rightarrow \sigma=0$, where $\avg{a^\dagger a} = \avg{b^\dagger b} = \sinh(\chi t)^2$.

\begin{figure}
    \centering
    \includegraphics[width=0.3\linewidth]{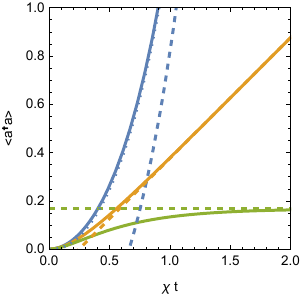}
    \caption{\textbf{Three regimes of TMS with symmetric losses.}
    Comparison between the exact prediction Eq.~\eqref{eqsupp:supp_ada_lossy} (solid lines) and the asymptotic behaviors Eq.~\eqref{eqsupp:ada_lossy_cases} (dashed lines), for $\sigma=0.05$ (blue), $\sigma=1$ (orange), and $\sigma=2$ (green).
    Note that, since we still have $\avg{a^\dagger a} \lesssim 1$ for the short evolution times considered, a deviation between the blue solid and dashed lines can be seen. This is mostly due to the replacement $e^{-\gamma t}\rightarrow 0$ in the last term of Eq.~\eqref{eqsupp:supp_ada_lossy}, which was not done for the blue dotted line, showing excellent agreement also in this short time scale.}
    \label{suppfig:evol_ada}
\end{figure}

Note that the parameter $\sigma$ defined above allows us to identify three different regimes, depending on whether $\sigma \gtreqless 1$. For $\sigma <1$ the number of excitations increases exponentially in time, for $\sigma =1$ it increases linearly, while for $\sigma >1$ it stays constant. The corresponding behavior is described by
\begin{equation}\label{eqsupp:ada_lossy_cases}
\avg{a^\dagger a} = 
\begin{cases}
     \frac{1}{\left(1-\sigma^2\right)} \left(e^{-\gamma t} (1 + \sigma) \sinh(\chi t)^2 - \frac{1}{2} \right) & \;\text{for }\;\sigma<1 \\
     \frac{1}{8} \left( 4 \chi t + e^{-4 \chi t}-1\right) & \;\text{for }\;\sigma=1 \\
     \frac{1}{2 \left(\sigma^2 - 1 \right)} & \;\text{for }\;\sigma>1 \\
\end{cases} \qquad\text{when }\; t\rightarrow\infty \;.
\end{equation}
As an illustration, these expressions are plotted in Fig.~\ref{suppfig:evol_ada}, together with Eq.~\eqref{eqsupp:supp_ada_lossy}, for three different values of $\sigma$.

\clearpage
\newpage

\section{Detection of continuous-variable entanglement and EPR steering}

\subsection{Entanglement criterion}

Entanglement between two bosonic modes can be revealed by the Giovannetti \etal criterion~\cite{giovannetti_characterizing_2003}. For the case of linear quadrature measurements satisfying $[X_i,P_j]=i\delta_{ij}$ it states that, for all separable states we have
\begin{equation}\label{eqsupp:giova}
    \var{X_a + g_x X_b} \var{P_a + g_p P_b} \geq \dfrac{1}{4}(1+\abs{g_x g_p})^2 \;,
\end{equation}
where $g_{x,p}$ are arbitrary real coefficients that can be freely adjusted in order to minimize the ratio between left- and right-hand sides of the inequality (i.e. in order to maximize its violation when possible).
Although a closed-form analytical expression for the optimal $g$'s in the general case is tedious, for a state with $\text{Var}[X_a]=\text{Var}[X_b]$ and $\text{Var}[P_a]=\text{Var}[P_b]$ one obtains the optimal gain parameters $g_x^\ast=-g_p^\ast=\pm 1$ (where the sign depends on whether the $X$ or $P$ quadratures are correlated.)
Using the arithmetic-geometric mean inequality $X+Y\geq 2 \sqrt{XY}$ (from $(\sqrt{X}-\sqrt{Y})^2\geq 0$), we obtain from Eq.~\eqref{eqsupp:giova} the separability criterion for the sum of variances
\begin{equation}\label{eqsupp:GiovSum}
    \Delta_\text{sum} = \var{X_a + g_x X_b} + \var{P_a + g_p P_b} \geq (1+\abs{g_x g_p}) \;.
\end{equation}
A violation of this inequality reveals entanglement.

\subsection{EPR steering criterion}

EPR correlations between two bosonic modes can be revealed by the Reid criterion~\cite{reid_demonstration_1989}, stating that for all states where B cannot steer A it holds
\begin{equation}\label{eqsupp:reid}
    \var{X_a + g_x X_b} \var{P_a + g_p P_b} \geq \dfrac{1}{4} \;.
\end{equation}
Note here that the optimal free parameters can be easily computed analytically, since they appear only on the left-hand side of the inequality. These are found as the coefficients minimizing each variance, namely as
\begin{equation}\label{eq:opt_g}
    \dfrac{\partial}{\partial g_x} \left( \var{X_a} + g_x^2 \var{X_b} + 2 g_x \cov{X_a}{X_b} \right) = 0 \qquad\Rightarrow\qquad g^\ast_x = - \dfrac{ \cov{X_a}{X_b} }{ \var{X_b} } \;,
\end{equation}
and similarly for $g_p$.
Using the inequality $X+Y\geq 2 \sqrt{X Y}$, we express criterion Eq.~\eqref{eqsupp:reid} in terms of the sum of variances as
\begin{equation}\label{eqsupp:ReidSum}
    \Delta_\text{sum} = \var{X_a + g_x X_b} + \var{P_a + g_p P_b} \geq 1 \;.
\end{equation}
A violation of this inequality reveals EPR steering from B to A, in symbols B$\rightarrow$A.

The directionality of steering becomes evident from the fact that, analogously to Eq.~\eqref{eqsupp:reid}, it is possible to derive the fact that for all states where A cannot steer B it holds
\begin{equation}\label{eqsupp:reidAB}
    \var{h_x X_a + X_b} \var{h_p P_a + P_b} \geq \dfrac{1}{4} \;.
\end{equation}
To clarify the difference, note that dividing the arguments of the variances in Eq.~\eqref{eqsupp:reid} for the corresponding $g$ and using $\var{cX}=c^2\var{X}$, we obtain $\var{X_a/g_x + X_b} \var{P_a/g_p + P_b} \geq 1/4 g_x^2 g_p^2$. 
This inequality is not equivalent to Eq.~\eqref{eqsupp:reidAB} through a simple relabeling $1/g\rightarrow h$, because the bound $1/4$ gets also rescaled.
Therefore, a violation of Eq.~\eqref{eqsupp:reid} does not necessarily imply a violation of Eq.~\eqref{eqsupp:reidAB} and vice versa. Quantum states violating only one of these two inequalities are said to show one-way steering, while states violating both of them show two-way steering.
Interestingly, this is different from the case of the entanglement criterion Eq.~\eqref{eqsupp:giova}, which is invariant under rescaling of the variances, consistent with the fact that entanglement is a symmetric form of quantum correlation.

\subsection{Witnessing quantum correlations from TMS interaction}

To characterize the quantum correlations generated by our effective TMS interaction, we would like to measure $\var{X_a + g_x X_b} + \var{P_a + g_p P_b}$, which requires local homodyne measurement of the quadratures in each output mode. This is currently difficult to achieve, as quadrature readout of the oscillator is not naturally implemented through a qubit, we therefore proceed with another strategy.

To derive our criterion for quantum correlations, let us start by using Eq.~\eqref{eq:abEOM1} to calculate $a_{\text{out}}^\dagger a_{\text{out}} + 1/2$, where $a_{\text{out}}$ refers to one of the output modes of TMS2 in Fig.~\ref{fig3}a of the main text. 
The result is
\begin{align}
    a_{\text{out}}^\dagger a_{\text{out}} + \dfrac{1}{2} &= (\cosh r \; a^\dagger + e^{-i\phi} \sinh r \; b)(\cosh r \; a + e^{i\phi} \sinh r \; b^\dagger) + \dfrac{1}{2} \notag\\
    &= (\cosh r \; a^\dagger + e^{-i\phi} \sinh r \; b)(\cosh r \; a + e^{i\phi} \sinh r \; b^\dagger) + \dfrac{1}{2} ( \cosh^2 r - \sinh^2 r ) \notag\\
    &= \cosh^2 r \left( a^\dagger a + \dfrac{1}{2} \right) + \sinh^2 r \left( b^\dagger b + \dfrac{1}{2} \right) + \cosh r \sinh r \left( e^{i\phi} a^\dagger b^\dagger + e^{-i\phi} a b \right) \;. \label{eq:spaghetti}
\end{align}
Here and throughout the following, we implicitly write $a_{\text{out}} \equiv a_{\text{out}}(\phi,r)$ to keep the notation compact. Moreover, let us remark that the operators $a$ and $b$ are the ones associated with the two modes between TMS1 and TMS2 in Fig.~\ref{fig3}a of the main text. These are the output of TMS1, where the state we want to characterize is prepared.

Introducing the mode quadratures as in Eqs.~\eqref{eq:quad0}, $X_a = (a+a^\dagger)/\sqrt{2}$ and $P_a = (a-a^\dagger)/i\sqrt{2}$, we obtain $X_a^2 + P_a^2 = 2 a^\dagger a + 1$, and similarly for $b$.
We can now invert Eqs.~\eqref{eq:quad2}, to obtain a set of equations for $a b$, $a b^\dagger$, $a^\dagger b$ and $a^\dagger b^\dagger$. We then use these to write the last term in Eq.~\eqref{eq:spaghetti} as
\begin{equation}\label{supp:adaFull}
    a_{\text{out}}^\dagger a_{\text{out}} + \dfrac{1}{2} = \cosh^2 r \left( \dfrac{X_a^2 + P_a^2}{2} \right) + \sinh^2 r \left( \dfrac{X_b^2 + P_b^2}{2} \right) + \cosh r \sinh r \left( \cos\phi (X_a X_b - P_a P_b) - \sin\phi (X_a P_b + P_a X_b) \right) \;.
\end{equation}
After dividing both sides by $(\cosh^2 r)/2$, we introduce the operator
\begin{align}
    \hat{\mathcal E}(\phi) &\equiv \dfrac{2 a_{\text{out}}^\dagger a_{\text{out}} + 1}{\cosh^2 r} \label{supp:adaFullbis} \\
    &= \left( X_a^2 + P_a^2 \right) + \tanh^2 r \left( X_b^2 + P_b^2 \right) + 2 \tanh r \left( \cos\phi (X_a X_b - P_a P_b) - \sin\phi (X_a P_b + P_a X_b) \right) \notag\\
    &= \bigl[X_a + \tanh r\, (\cos \phi X_b - \sin\phi P_b) \bigr]^2 + \bigl[P_a - \tanh r\,(\sin\phi X_b + \cos\phi P_b)\bigr]^2 \notag\;.
\end{align}
Recalling the definition of quadrature vector $\bm{\xi}=(X_a,P_a,X_b,P_b)^T$ and covariance matrix $V_{ij} = \frac{1}{2}\avg{\xi_i \xi_j + \xi_j \xi_i} - \avg{\xi_i}\avg{\xi_j}$, we can write
\begin{equation}\label{supp:eqEphi}
    \avg{\hat{\mathcal E}(\phi)} = u^T V u + w^T V w + (u^T\avg{\bm{\xi}})^2 + (w^T\avg{\bm{\xi}})^2 
\end{equation}
where we have introduced the vectors
\begin{equation}
u =
\begin{pmatrix}
1\\
0\\
\tanh r\cos\phi\\
-\tanh r\sin\phi
\end{pmatrix},
\qquad
w =
\begin{pmatrix}
0\\
1\\
-\tanh r\sin\phi\\
-\tanh r\cos\phi
\end{pmatrix} \;.
\label{eq:u_w_vectors}
\end{equation}
A simpler expression is obtained by defining the rotated quadratures 
\begin{equation}
    X_b(\phi) = \cos \phi X_b - \sin\phi P_b \;,\qquad P_b(\phi) = \sin\phi X_b + \cos\phi P_b \;,
\end{equation}
such that $u^T V u
= \mathrm{Var}[u^T \bm{\xi}]
= \mathrm{Var}\left[X_a+\tanh r\,X_b(\phi)\right]$ and $u^T\bm{\xi}
= X_a+\tanh r(\cos\phi\,X_b-\sin\phi\,P_b)
= X_a+\tanh r\,X_b(\phi)$, and similarly for the terms with $w$. 
Combining all terms allows us to obtain from Eq.~\eqref{supp:eqEphi} the inequality
\begin{equation}\label{eqsupp:derivVVsum} 
    \avg{ \hat{\mathcal E}(\phi) } \geq \text{Var}\left[ X_a + \tanh r X_b(\phi) \right] +  \text{Var}\left[ P_a - \tanh r P_b(\phi) \right] \;,
\end{equation}
with the inequality saturated if and only if both $\avg{X_a + \tanh r X_b(\phi)}=0$ and $\avg{P_a - \tanh r P_b(\phi)}=0$, such as for states with zero first moments. This gives the expression for $\mathcal{E}(\phi)=\avg{\mathcal{\hat E}(\phi)}$
\begin{align}\label{supp:crit}
    \mathcal{E}(\phi) \equiv \dfrac{2 \langle a_{\text{out}}^\dagger a_{\text{out}} \rangle + 1}{\cosh^2 r} &\geq  \text{Var}\left[ X_a + g_x X_b(\phi) \right] +  \text{Var}\left[ P_a + g_p P_b(\phi) \right] \\
    g_x &= \tanh r = -g_p
\end{align}
which we presented in the main text.
Finally, looking at Eqs.~(\ref{eqsupp:GiovSum},\ref{eqsupp:ReidSum}), we can conclude that
\begin{align}
    \text{for all separable states}\qquad &\mathcal{E}(\phi) \geq ( 1 + \tanh^2 r ) \geq 1 \\
    \text{for all non-steerable B$\rightarrow$A states}\qquad &\mathcal{E}(\phi) \geq 1 \;. \label{suppeq:EPRcritBA}
\end{align}
Thus, a violation of these inequalities for some $\phi$ reveals entanglement or EPR steering B$\rightarrow$A, respectively.

In order to compute $\mathcal{E}(\phi)$ we have to access two quantities. The first is $\langle a_{\text{out}}^\dagger a_{\text{out}} \rangle$, namely the average number of excitations at the output of the TMS2 operation when the two-mode state under investigation is fed as input. 
The second is $\cosh^2 r = 1 + \overline{N_a}$, namely the average number of excitations resulting from the same TMS2 operation when two-mode vacuum is fed as an input. The latter can be understood as a reference measurement of the ``shot noise'' level to be surpassed.

Note that the gain parameters $g_{x,p}$, as well as the phase $\phi$, can be adjusted by deciding the strength $r$ and phase of the TMS2 operation. Remember that there is an optimal value for the gain parameters, which depends on the criterion under investigation as discussed in the previous sections.
For detecting EPR steering B$\rightarrow$A on a TMS vacuum state generated by a TMS1 operation with strength $r_1$, the covariance matrix Eq.~\eqref{eq:cm_tms} allows us to compute the optimal parameter Eq.~\eqref{eq:opt_g}, giving $g_x^\text{opt} = - \tanh (2 r_1) = - g_p^\text{opt}$. Interestingly, this means that the optimal TMS2 strength for readout should be twice larger than the one for state preparation. For this choice, and at the optimal phase $\phi=\pi$, the criterion \eqref{supp:crit} reads $\mathcal{E}=1/\cosh(2 r_1)$, which gives $\mathcal{E}<1$ for any $r_1 > 0$.

Notably, EPR steering is an asymmetric form of quantum correlations, meaning that a state for which B can steer A, B$\rightarrow$A, not necessarily implies that also A can steer B, A$\rightarrow$B. These two steering directions are detected by Eqs.~\eqref{eqsupp:reid} and \eqref{eqsupp:reidAB}, respectively. Above, we have shown the derivation of criterion Eq.~\eqref{suppeq:EPRcritBA} for B$\rightarrow$A steering, but a similar criterion for A$\rightarrow$B steering follows straightforwardly by defining instead of Eq.~\eqref{supp:adaFullbis} the operator
\begin{equation}
    \hat{\mathcal E}(\phi) \equiv \dfrac{2 b_{\text{out}}^\dagger b_{\text{out}} + 1}{\cosh^2 r} \;. \label{supp:adaFullbisAB}
\end{equation}
This is naturally implemented by just measuring the other output mode of TMS2. In the main text, we show the experimental violation of both criteria for B$\rightarrow$A and A$\rightarrow$B steering.

\clearpage
\newpage

\section{Properties of two-mode Gaussian states}

\subsection{Entanglement condition}

Given the quadrature vector $\bm{\xi}=(X_a,P_a,X_b,P_b)^T$, Gaussian states are fully characterized by the first moments $\avg{\bm{\xi}}$ and by the covariance matrix $V_{ij} = \frac{1}{2}\avg{\xi_i \xi_j + \xi_j \xi_i} - \avg{\xi_i}\avg{\xi_j}$.
For any two-mode Gaussian state (mixed or pure), one can always choose local Gaussian unitary transformations such that first moments are zero and the CM takes the standard form
\begin{equation} \label{eq:cm_std}
    V =  \left( \begin{matrix}
v_1 & 0 &  c_1 &  0 \\
0 & v_1 &  0 & c_2 \\
 c_1 & 0 & v_2 & 0 \\
0 & c_2 & 0 & v_2 
\end{matrix}\right) ,
\end{equation}
with $v_1,v_2\geq 1/2$ representing local variances of the modes and $c_1,c_2\in\mathbb{R}$ inter-mode correlations.
For example, the covariance matrix of a TMSV state Eq.~\eqref{eq:cm_tms} can be written in this form for $\phi=0$, which gives $v_1=v_2=\cosh(2r)/2$ and $c_1=-c_2=\sinh(2r)/2$.

For two--mode Gaussian states, separability is completely characterized by the positivity of the partial transpose (PPT) criterion. In the covariance-matrix formalism, partial transposition corresponds to a mirror reflection of one momentum quadrature, e.g. $P_b \to -P_b$. This is implemented as
\[
V^\Gamma = \Lambda V \Lambda,
\qquad
\Lambda = \mathrm{diag}(1,1,1,-1),
\]
which for the case of Eq.~\eqref{eq:cm_std} yields
\begin{equation}\label{eq:cm_ppt}
V^\Gamma =
\begin{pmatrix}
v_1 & 0   & c_1 & 0    \\
0   & v_1 & 0   & -c_2 \\
c_1 & 0   & v_2 & 0    \\
0   & -c_2& 0   & v_2
\end{pmatrix}.
\end{equation}

To evaluate the PPT entanglement condition, we consider the following.
For a generic two-mode covariance matrix written in block form
\(
V=\big(\begin{smallmatrix}A & C\\ C^T & B\end{smallmatrix}\big)
\),
we introduce the symplectic invariants 
\begin{align}
    \det V &= (v_1 v_2 - c_1^2)(v_1 v_2 - c_2^2) \\
    \Delta &= \det A + \det B + 2 \det C = v_1^2 + v_2^2 + 2c_1 c_2 \;.
\end{align}
From these, the symplectic eigenvalues of $V$ are given by (note the square on the left hand side)
\[
\nu_{\pm}^2
= \frac{1}{2}\left(
\Delta \pm \sqrt{\Delta^2 - 4 \det V}
\right) .
\]
In the chosen units, the uncertainty principle requires $\nu_{\pm} \ge 1/2$. 
According to the PPT criterion, a two-mode Gaussian state is separable if and only if the partially transposed covariance matrix also satisfies the uncertainty principle. 
Therefore, the state is entangled if and only if the smallest symplectic eigenvalue of $V^\Gamma$ violates this bound, i.e. $\tilde\nu_- < 1/2$ implies entanglement.
Concretely, this means that
\begin{equation}\label{supp:eqPPT}
    \tilde\nu_-^2 = \frac{1}{2}\left(
\tilde\Delta
- \sqrt{\tilde\Delta^2 - 4\det \tilde V}
\right)
< \frac{1}{4}
\end{equation}
provides a necessary and sufficient condition for entanglement.
From Eq.~\eqref{eq:cm_ppt} we have $\tilde\Delta = v_1^2 + v_2^2 - 2 c_1 c_2$ and $\det \tilde V = (v_1 v_2 - c_1^2)(v_1 v_2 - c_2^2)$. If $v_1=v_2=v$ and $c_1=-c_2=c$ criterion \eqref{supp:eqPPT} becomes $(v-|c|)<1/2$, meaning that when considering Eq.~\eqref{eq:cm_tms} we have that any $r>0$ implies entanglement, as expected from an ideal TMSV state.

\subsection{Logarithmic negativity from the PPT criterion}

A widely used quantitative measure of entanglement for continuous--variable systems is the \emph{logarithmic negativity}, which is directly based on the positivity of the partial transpose (PPT) criterion. For a bipartite quantum state \(\rho\), it is defined as
\[
E_N \equiv \log_2 \|\rho^\Gamma\|_1,
\]
where \(\rho^\Gamma\) denotes the partially transposed density operator and \(\|\cdot\|_1\) is the trace norm.

For two-mode Gaussian states, the logarithmic negativity can be computed entirely at the level of the covariance matrix by taking the minimum symplectic eigenvalue of its partial transpose $\tilde \nu_-$. It is given by
\begin{equation} \label{supp:defEN}
    E_N =
\max\left[
0,\;
-\log_2\left(2\,\tilde\nu_-\right)
\right] \;.
\end{equation}
If \(\tilde\nu_- \ge 1/2\), the partial transpose is physical and the logarithmic negativity vanishes. If \(\tilde\nu_- < 1/2\), the PPT bound is violated and the state is entangled, with \(E_N\) quantifying the degree of this violation.

\subsection{Entanglement quantification from $\mathcal{E}$}

We show here that, for Gaussian states, the experimentally accessible quantity $\mathcal{E}$ defined in Eq.~\eqref{supp:crit} provides an upper bound on the smallest symplectic eigenvalue \(\tilde\nu_-\) of the partially transposed covariance matrix \(V^\Gamma\), and therefore a lower bound on the logarithmic negativity. 
Note that this is not immediately obvious, since local Gaussian transformations can change $\mathcal{E}$ arbitrarily for fixed gains $g_x$, $g_p$, even though $\tilde\nu_-$ and $E_N$ are invariant.
However, the ratio $\mathcal{E}/(1+|g_x g_p|)$ cannot be arbitrarily small, thus allowing us to derive nontrivial bounds on $E_N$.

Let us consider an experimentally measured value $\mathcal E^{\exp}$ for gains $g_x,g_p$ and introduce the normalized quantity
\begin{equation}
\epsilon_{\exp} = \frac{\mathcal E^{\exp}}{1+|g_x g_p|}.
\label{eq:epsilon_exp}
\end{equation}
From Eq.~\eqref{eqsupp:GiovSum} we have that all separable states satisfy $\epsilon_{\exp}\geq 1$, so that \(\epsilon_{\exp}<1\) certifies entanglement.
We now show that, for any two-mode Gaussian state, the observed value \(\epsilon_{\exp}\) provides a rigorous lower bound on the logarithmic negativity.
The key object is the smallest symplectic eigenvalue \(\tilde\nu_-\) of the partially transposed covariance matrix \(V^\Gamma\), which fully characterizes entanglement for Gaussian states via the PPT criterion. In the present convention, the logarithmic negativity is given by Eq.~\eqref{supp:defEN}.
To relate \(\epsilon_{\exp}\) to \(\tilde\nu_-\), define the globally optimized quantity
\begin{equation}
\epsilon_\star
\equiv
\inf_{S_a\oplus S_b}
\;
\inf_{g_x,g_p\in\mathbb{R}}
\frac{\mathcal E (g_x,g_p)}{1+|g_x g_p|},
\label{eq:epsilon_star}
\end{equation}
where $\mathcal E (g_x,g_p)$ is evaluated on the covariance matrix
\(V_S=(S_a\oplus S_b)V(S_a\oplus S_b)^T\), and the infimum is taken over all local
symplectic transformations \(S_a\oplus S_b\).
Local symplectics correspond to local Gaussian unitaries and therefore do not
change \(\tilde\nu_-\) or \(E_N\).
By definition, for any fixed basis and any fixed gains (in particular, those used
experimentally),
\begin{equation}
\epsilon_\star \le \epsilon_{\exp}.
\label{eq:star_le_exp}
\end{equation}

Because the infimum in Eq.~\eqref{eq:epsilon_star} allows for arbitrary local symplectic transformations, we may evaluate \(\epsilon_\star\) in the symmetric standard form of the covariance matrix,
\begin{equation}\label{eq:cm_sym}
V =
\begin{pmatrix}
v & 0 & c & 0 \\
0 & v & 0 & -c \\
c & 0 & v & 0 \\
0 & -c & 0 & v
\end{pmatrix},
\qquad v\ge \tfrac12,\; c\in\mathbb{R}.
\end{equation}
For this form, the smallest symplectic eigenvalue of the partially transposed covariance matrix is $\tilde\nu_- = v-|c|$.
Consider gains of the form $g_x=g=-g_p$.
A direct calculation gives
\begin{equation}
\mathcal E = 2v(1+g^2)+4cg, \qquad\Rightarrow\qquad \frac{\mathcal E}{1+g^2}
=
2v+\frac{4cg}{1+g^2} \;.
\end{equation}
Optimizing over \(g\) yields a minimum at \(g=-\text{sign}(c)=\pm 1\) (for the entangled case \(c\neq 0\) and we can choose $c>0$ without loss of generality to have $g=1$), giving
\begin{equation}
\epsilon_{\mathrm{opt}}^{(\mathrm{sym})}
=
2v+2c
=
2(v-|c|)
=
2\tilde\nu_-.
\label{eq:epsilon_opt_sym}
\end{equation}
Since this value is achieved after optimizing both the local symplectic basis and
the gains, it follows that
\begin{equation}
\epsilon_\star = 2\tilde\nu_-.
\label{eq:epsilon_star_nu}
\end{equation}

Combining \eqref{eq:star_le_exp} and \eqref{eq:epsilon_star_nu}, we obtain the bound
\begin{equation}
\tilde\nu_- \le \frac{\epsilon_{\exp}}{2}.
\label{eq:nu_bound}
\end{equation}
Substituting this inequality into the definition
\eqref{supp:defEN} of the logarithmic negativity gives
\begin{equation}
E_N
\ge
\max\bigl\{0,\,-\log_2(\epsilon_{\exp})\bigr\}.
\label{eq:EN_bound}
\end{equation}

Thus, for any two-mode Gaussian state, an experimentally observed value of the
normalized ratio \(\epsilon_{\exp}\) at arbitrary gains provides a rigorous
lower bound on the logarithmic negativity.
The bound is tight for symmetric standard Gaussian states with optimally chosen
gains, and becomes conservative when the experimental basis or gains are not
optimal.

\clearpage
\newpage

\vspace{30mm}
\section{SU(1,1) interferometry and quantum-enhanced metrology}

\subsection{The SU(1,1) interferometer}

While traditional Mach-Zehnder-type interferometers use passive optical elements such as beam splitters, SU(1,1) interferometers replace them with active optical elements, in particular with parametric amplifiers.
Their name originates from the fact that, while the passive mode operations in a Mach-Zehnder interferometer can be understood as elements of the SU(2) algebra, parametric amplifiers result in transformations obeying the SU(1,1) algebra. In fact, introducing
\begin{equation}
    K_x = \dfrac{1}{2} (a^\dagger b^\dagger + a b ) \;,\qquad
    K_y = \dfrac{1}{2i} (a^\dagger b^\dagger - a b ) \;,\qquad
    K_z = \dfrac{1}{2} (a^\dagger a + b^\dagger b + 1 )
\end{equation}
we have that these operators satisfy the commutation relations
\begin{equation}\label{supp:su11comm}
    [K_x,K_y] = - i K_z \;,\qquad [K_y,K_z] = i K_x \;,\qquad [K_z,K_x] = i K_y \;,
\end{equation}
that define the SU(1,1) group. Note that $K_{x,y}$ describe the coherent and pairwise creation and destruction of excitations, while $K_z$ is related to the total number of excitations in the two modes. 
The Casimir invariant $K_{\text{tot}}^2 = K_z^2 - K_x^2 - K_y^2 = \frac{1}{4}((a^\dagger a - b^\dagger b)^2 - 1)$ reflects the fact that the imbalance between the two modes is constant.

Geometrically, while mode operations in a Mach-Zehnder interferometer can be illustrated as rotations on a (Bloch) sphere, operations in an SU(1,1) interferometer can be illustrated as boosts (for $K_{x,y}$) and rotations (for $K_z$) on the hyperbolic surface defined by $K_{\text{tot}}^2$. 
The bottom point of this surface is the vacuum state $\ket{0,0}$ and, since the total number of excitations is in principle unbounded, this surface is open on the top. 

The action of $K_{x,y}$ as boosts can be seen from writing the TMS operator Eq.~\eqref{eq:TMSopDef} as
\begin{align}
  S(r,\phi) &= \exp\bigl[r e^{i\phi} a^\dagger b^\dagger - r e^{-i\phi} a b\bigr] \\
  &= \exp\Bigl[2 i r\bigl(\sin\phi\, K_x + \cos\phi\, K_y\bigr)\Bigr],
\end{align}
which realizes the Bogoliubov transformations Eqs.~\eqref{eq:abEOM1}. On the other hand, the action of $K_z$ as a phase-shift can be seen from the fact that the unitary
\begin{equation}
  U_\theta = \exp \left[ -i\theta K_z \right],
\end{equation}
realizes the transformations $U_\theta^\dagger\, a\, U_\theta = e^{-i\theta/2} a$ and $U_\theta^\dagger\, b\, U_\theta = e^{-i\theta/2} b$.
To summarize, the evolution of the bosonic operators $v=(a,b^\dagger)^T$ in the Heisenberg picture is given by
\begin{equation}\label{supp:heis}
    M_S(r,\phi) = 
  \begin{pmatrix}
    \cosh r & e^{i\phi} \sinh r\\
    e^{-i\phi} \sinh r & \cosh r
  \end{pmatrix}, \qquad
    M_\theta =
  \begin{pmatrix}
    e^{-i\theta/2} & 0\\
    0 & e^{i\theta/2}
  \end{pmatrix},
\end{equation}
for the $S(r,\phi)$ and $U_\theta$ operators, respectively.
Crucially, given the commutation relations \eqref{supp:su11comm}, we have the identity
\begin{equation}\label{supp:eqRot}
  \exp\Bigl[2 i r\bigl(\sin\phi\, K_x + \cos\phi\, K_y\bigr)\Bigr] \exp[-i\theta K_z]
  =
  \exp[-i\theta K_z]\,
  \exp\Bigl[2ir(\sin(\phi+\theta)\,K_x+\cos(\phi+\theta)\,K_y)\Bigr].
\end{equation}
This is obtained from
\(e^{A}e^{B}=e^{B
}(e^{-B} e^{A} e^{B})\), with
\(A=2ir(\sin\phi\,K_x+\cos\phi\,K_y)\), \(B=-i\theta K_z\), and
\begin{equation}
e^{-B} e^{A} e^{B} = e^{-B}\Bigl(\sum_{n=0}^\infty \frac{A^n}{n!}\Bigr)e^{B} = \sum_{n=0}^\infty \frac{e^{-B} A^n e^{B}}{n!} = \sum_{n=0}^\infty \frac{(A')^n}{n!} = e^{A'},
\end{equation}
where $A' = e^{-B} A e^{B} = A - [B,A] + \frac{1}{2!}[B,[B,A]] - \frac{1}{3!}[B,[B,[B,A]]] + \dots$ 
Inserting $[B,K_x]=\theta K_y$, $[B,K_y]=-\,\theta K_x$ and $[B,[B,A]]=-\theta^2 A$ finally gives Eq.~\eqref{supp:eqRot}.
This result allows us to understand that rotating the phase of the second TMS operation has the same effect as adding a phase shift within the interferometer, since the last rotation is irrelevant if one performs number-counting measurement at the outputs.

The SU(1,1) interferometric sequence consists of two boosts with a phase rotation in-between. Starting from the two-mode initial state $\ket{\psi_\text{in}}$, it corresponds to the transformation
\begin{align}
    \ket{\psi_\text{out}} &= U_{\mathrm{SU}(1,1)}(r_2,\phi,\theta,r_1) \ket{\psi_\text{in}} \\
    &= S(r_2,\phi)\, U_\theta \, S(r_1,0) \ket{\psi_\text{in}} \;.
\end{align}
This is easily computed using Eqs.~\eqref{supp:heis} as
\begin{equation}
  \begin{pmatrix} a_{\mathrm{out}} \\ b_{\mathrm{out}}^\dagger \end{pmatrix}
  =
  M_S(r_2,\phi) \, M_\theta\, M_S(r_1,0)
  \begin{pmatrix} a \\ b^\dagger \end{pmatrix}.
\end{equation}
The output modes take the form
\begin{align}\label{supp:BogSU11}
a_{\mathrm{out}} &=
A(r_1,r_2,\phi,\theta)\, a
+ B(r_1,r_2,\phi,\theta)\, b^\dagger, \\
b_{\mathrm{out}}^\dagger &=
B^*(r_1,r_2,\phi,\theta)\, a
+ A^*(r_1,r_2,\phi,\theta)\, b^\dagger,
\end{align}
where we have introduced the coefficient functions
\begin{align}
A(r_1,r_2,\phi,\theta)
&\equiv
e^{-i\theta/2} \left[ \cosh r_1 \cosh r_2
+ e^{i(\phi+\theta)} \sinh r_1 \sinh r_2 \right],
\\
B(r_1,r_2,\phi,\theta)
&\equiv
e^{-i\theta/2} \left[ \sinh r_1 \cosh r_2
+ e^{i(\phi+\theta)} \cosh r_1 \sinh r_2 \right] ,
\end{align}
satisfying \( |A|^2 - |B|^2 = 1 \), as required for an SU(1,1) evolution.

Typically, what is measured is the intensity at the output ports of the interferometer. Taking as an example mode $a$, this intensity is computed as the expectation value of the phonon number operator $N_a^{\mathrm{(out)}}(\theta) = a_{\mathrm{out}}^\dagger(\theta)\, a_{\mathrm{out}}(\theta)$.
Using the Bogoliubov form Eq.~\eqref{supp:BogSU11} and computing the expectation value in the input vacuum $\ket{0,0}$, we obtain
\begin{align}
  \bigl\langle N_a^{\mathrm{(out)}}(\theta)\bigr\rangle
  &=  |B|^2 = \sinh^2 r_1 \cosh^2 r_2 + \sinh^2 r_2 \cosh^2 r_1 + 2 \sinh r_1 \cosh r_2 \sinh r_2 \cosh r_1 \cos(\phi+\theta) \\
  \bigl\langle (N_a^{\mathrm{(out)}}(\theta))^2 \bigr\rangle &= 2|B|^4 + |B|^2  \;.
\end{align}
The mean output population is thus a sinusoidal function of the phase $\theta$.
As expected from symmetry considerations, the same result is obtained for the first and second moments of $N_b^{\mathrm{(out)}}(\theta)$. 
For the total phonon number we then have
\begin{align}
    \bigl\langle N_a^{\mathrm{(out)}}(\theta) + N_b^{\mathrm{(out)}}(\theta)\bigr\rangle &= 2 |B|^2 \\
    \bigl\langle ( N_a^{\mathrm{(out)}}(\theta) + N_b^{\mathrm{(out)}}(\theta) )^2 \bigr\rangle &= 8 |B|^4 + 4 |B|^2  \;. \label{suppeq:Nmeanvar}
\end{align}
These results will be used in the calculation of the classical Fisher information from phonon number measurements.

\subsection{Phase sensitivity and quantum enhancement}

We now evaluate the quantum Fisher information (QFI) for estimating the phase $\theta$ in the above $\mathrm{SU}(1,1)$ interferometer.  
Considering a phase imprinting generated by a unitary operator on a pure TMSV state, the QFI is 
\begin{equation}\label{supp:QFItmsv}
  F_Q = 4\,\var{K_z}_{\ket{\psi_\text{TMSV}}} = \sinh^2(2r_1) \;.
\end{equation}
As the total mean phonon number is $\overline{N}=\overline{N_a}+\overline{N_b}=2\sinh^2 r_1$, with $\overline{N_a}=\overline{N_b}$, we have $F_Q =4 \sinh^2 r_1 \cosh^2 r_1 =\overline{N}(\overline{N}+2)=4\overline{N_a}(\overline{N_a}+1)$, showing that the QFI displays Heisenberg scaling $F_Q\sim\bar{N}^2$ at large excitation number. The shot-noise limit is instead $F_Q=\overline{N}=2 \overline{N_a}$, meaning that observing a Fisher information larger than this value implies nonclassical states and/or correlations.

It is also interesting to derive this result from the point of view of local phase shifts on the two bosonic modes, using the formalism of the QFI matrix (QFIM).
These local phase shifts are given by $U_a(\theta_a)=e^{-i \theta_a \hat N_a}$ and similarly for $b$.
For a pure state, the QFIM is given by
\begin{equation}
F_{ij}
=
4\,\mathrm{Cov}(\hat G_i,\hat G_j),
\end{equation}
where $G_{i,j}\in\{\hat N_a, \hat N_b\}$ are the generators of local phase shifts.
For a TMSV state, variances and covariances of these operators have been calculated in Eqs.~(\ref{suppeq:varNa},\ref{suppeq:covNaNb}), giving $\mathrm{Var}(\hat N_a)=\mathrm{Var}(\hat N_b) = \overline{N_a}(\overline{N_a}+1)$ and $\mathrm{Cov}(\hat N_a,\hat N_b) = \overline{N_a}(\overline{N_a}+1)$, with $\overline{N_a}=\sinh^2 r_1$.
The QFIM therefore takes the form
\begin{equation}
F = 4\overline{N_a}(\overline{N_a}+1)
\begin{pmatrix}
1 & 1\\
1 & 1
\end{pmatrix}.
\end{equation}
For an arbitrary linear combination of phases $\theta = u_a\theta_a+u_b\theta_b$ generated by $\hat G_\theta=u_a\hat N_a+u_b\hat N_b$, the corresponding QFI is
\begin{equation}
F_Q(\theta)=\boldsymbol{u}^\mathsf{T}F\boldsymbol{u},
\qquad
\boldsymbol{u}=(u_a,u_b)^\mathsf{T}.
\end{equation}
The matrix \(F\) has eigenvalues
\(
\lambda_+=8\overline{N_a}(\overline{N_a}+1)
\)
and
\(
\lambda_-=0
\),
with normalized eigenvectors $\boldsymbol{v}_\pm=(1,\pm1)/\sqrt{2}$.
Hence, the optimal phase combination is the common phase
\begin{equation}
\theta_+=\frac{\theta_a+\theta_b}{2},
\qquad
\hat G_{+}=\frac{\hat N_a+\hat N_b}{2} \propto \hat K_z,
\end{equation}
for which the QFI is $F_Q = \sinh^2(2r_1)=4\overline{N_a}(\overline{N_a}+1)$, as found from Eq.~\eqref{supp:QFItmsv}.
The phase difference $\theta_-=(\theta_a-\theta_b)/2$ has vanishing QFI, reflecting the invariance of the TMSV under opposite local phase shifts.

The crucial question we are left now to analyze is whether this fundamental precision bound given by the QFI can be saturated by measurements available in our experiment, namely if the classical Fisher information associated with the probability distribution of the measurement results coincides with the quantum Fisher information just obtained.

\vspace{3mm}
\textbf{Mean phonon number measurement. --} According to the so-called method of moment, the sensitivity to a parameter $\theta$ when measuring a signal $N(\theta)$ is given by
\begin{equation}\label{suppeq:errpropChi}
    \chi(\theta) = \dfrac{\var{N(\theta)}}{|\partial N(\theta)/\partial\theta|^2} \;.
\end{equation}
Considering the result Eq.~\eqref{suppeq:Nmeanvar}, this is
\begin{equation}\label{supp:chiSU11}
\chi(\theta)
= \frac{\operatorname{Var}[N(\theta)]}{\left|\partial_\theta N(\theta)\right|^2}
= \frac{4|B(r_1,r_2,\phi,\theta)|^2\big(|B(r_1,r_2,\phi,\theta)|^2+1\big)}
       {16\,K^2\sin^2(\phi+\theta)}
= \frac{|B|^2\big(|B|^2+1\big)}{4K^2\sin^2(\phi+\theta)},
\end{equation}
where remember we have
\begin{equation}
|B|^2
= s_1^2 c_2^2 + c_1^2 s_2^2 + 2K\cos(\phi+\theta),\qquad
K = s_1 c_2 s_2 c_1,\quad
s_j=\sinh r_j,\; c_j=\cosh r_j.
\end{equation}

Interestingly, in contrast to the SU(2) case, the same sensitivity is achieved by measuring a single interferometer output, e.g. $N_a=\ada$, as by jointly measuring both outputs, $N=\ada+\bdb$.
This can change in a realistic scenario where noise is present.

In the special case of a symmetric SU(1,1) interferometer we have $r_1=r_2=r$, taking $\phi=0$ gives $K=\sinh^2(2r)/4$ and $|B|^2=4 K \cos^2(\theta/2)$. Therefore we arrive at $\chi^{-1}(\theta)=\frac{4K\sin^2(\theta/2)}{4K\cos^2(\theta/2)+1}$, which is maximized at $\theta=\pi$, where $\chi^{-1}(\pi)=\sinh^2(2r)=F_Q$.
This shows that the QFI can be saturated by the considered measurement.

Another interesting scenario is the high-gain amplifier limit where $r_2 \gg r_1$. In this case, fixing $\phi=0$ for simplicity, we find $\chi^{-1}(\theta)= \frac{\sinh^2(2r_1)\sin^2\theta}{(\cosh(2r_1)+\sinh(2 r_1)\cos\theta)^2}$, showing that the sensitivity becomes independent of \(r_2\). 
Maximizing this expression with respect to $\theta$ yields the condition $\cos\theta^\star=-\tanh(2 r_1)$, which has two solutions in \([0,2\pi)\), namely $\theta^\ast = \pi \pm \arccos(\tanh(2r_1))$.
At these points we find again $\chi^{-1}(\theta^\ast)=\sinh^2(2r_1)=F_Q$.

Considering the data in Fig.~\ref{fig3}d of the main text, note that before the action of TMS2 we have $\overline{N_b}\simeq 0.25$. Assuming the state is an ideal TMSV state, this would correspond to a QFI of $F_Q^{\rm TMSV}=4\overline{N_b}(\overline{N_b}+1)\simeq 1.25$. On the other hand, assuming $\overline{N_a}\approx\overline{N_b}$, the SQL would be $F_Q^{\rm SQL}=2\overline{N_b}\approx 0.5$.
We can now compare these with the classical FI obtained from the interferometer fringe Fig.~\ref{fig3}e.
A fit of the fringe to the model $\overline{N_b}=A\cos(\phi+\phi_0)+C$ gives $A\simeq 0.327$, $\phi_0\simeq -0.114$, and $C\simeq 0.515$. Taking the derivative with respect to $\phi$ and recalling that for a thermal distribution $\operatorname{Var}[\overline{N_b}(\phi)]=\overline{N_b}(\phi)(\overline{N_b}(\phi)+1)$, we obtain $\min_\phi \chi(\phi)\simeq 5.9$, i.e. $F^{\rm exp}\simeq 0.17$. 

We note that a value $F^{\rm exp}$ below $F_Q^{\rm TMSV}$ and $F_Q^{\rm SQL}$ is not necessarily surprising, especially by looking at the measured fringe in Fig.~\ref{fig3}e. 
An ideal SU(1,1) interferometer has maximum sensitivity at the dark fringe, namely around the phase $\phi'$ where $\overline{N_b}=0$. This is because, although a change of $\overline{N_b} \sim \sin^2(\phi-\phi')$ occurs there only to second order in $\phi$, the uncertainty $\operatorname{Var}[\overline{N_b}]=\overline{N_b}(\overline{N_b}+1)$ has also (to lowest order) a second-order dependence.
For finite losses, or asymmetric interferometers with $r_1\neq r_2$, TMS2 can never achieve $\overline{N_b}=0$, thus resulting in a finite background offset which for us amounts to $C-A\simeq 0.188$. For this background level, $F^{\rm exp} > F_Q^{\rm SQL}$ can only be achieved if $A\gtrsim 0.75$.

\vspace{3mm}
\textbf{Phonon distribution measurement. --} One could wonder whether full counting statistics of one or both output ports of the interferometer could allow for saturating the QFI for any $\theta$.
Direct calculation gives for one mode $p(n|\theta)=\frac{1}{N(\theta)+1}
\left(\frac{N(\theta)}{N(\theta)+1}\right)^n$, with $N(\theta)=|B(r_1,r_2,\phi,\theta)|^2$, and for two modes $p(N_a,N_b|\theta)=p(N_a|\theta)\,\delta_{N_a,N_b}$, showing that both single-mode and two-mode counting statistics are fully specified by the same scalar parameter \(N(\theta)\).
As a result, the score function \(\partial_\theta\ln p(n|\theta)\) is linear in the measurement outcome \(n\), implying that the classical Fisher information is completely exhausted by the first moment. Explicitly,
\begin{equation}
F_C^{\hat N}(\theta)
=
\sum_n \frac{[\partial_\theta p(n|\theta)]^2}{p(n|\theta)}
=
\frac{[\partial_\theta N(\theta)]^2}{N(\theta)\big(N(\theta)+1\big)},
\end{equation}
and no additional information is contained in higher moments or in joint photon-number correlations between the two modes. 
Importantly, note here that this expression coincides with $\chi^{-1}$ as defined in Eq.~\eqref{suppeq:errpropChi}.

Therefore, full photon counting, either on a single output mode or jointly on both modes, cannot provide more phase information than that already accessible from the mean photon number. Any gap between the resulting classical Fisher information and the quantum Fisher information is not due to incomplete use of the counting statistics, but rather to the choice of measurement itself: photon counting in the output Fock basis is not, in general, the optimal POVM for estimating the phase generated by \(K_z\).

\end{document}